# Machine learning models predict calculation outcomes with the transferability necessary for computational catalysis

Chenru Duan[1,2], Aditya Nandy[1,2], Husain Adamji[1], Yuriy Roman-Leshkov[1], and Heather J. Kulik[1,*]

[1]*Department of Chemical Engineering, Massachusetts Institute of Technology, Cambridge, MA 02139*

[2]*Department of Chemistry, Massachusetts Institute of Technology, Cambridge, MA 02139*

ABSTRACT: Virtual high throughput screening (VHTS) and machine learning (ML) have greatly accelerated the design of single-site transition-metal catalysts. VHTS of catalysts, however, is often accompanied with high calculation failure rate and wasted computational resources due to the difficulty of simultaneously converging all mechanistically relevant reactive intermediates to expected geometries and electronic states. We demonstrate a dynamic classifier approach, i.e., a convolutional neural network that monitors geometry optimization on the fly, and exploit its good performance and transferability for catalyst design. We show that the dynamic classifier performs well on all reactive intermediates in the representative catalytic cycle of the radical rebound mechanism for methane-to-methanol despite being trained on only one reactive intermediate. The dynamic classifier also generalizes to chemically distinct intermediates and metal centers absent from the training data without loss of accuracy or model confidence. We rationalize this superior model transferability to the use of on-the-fly electronic structure and geometric information generated from density functional theory calculations and the convolutional layer in the dynamic classifier. Combined with model uncertainty quantification, the dynamic classifier saves more than half of the computational resources that would have been wasted on unsuccessful calculations for all reactive intermediates being considered.

## 1. Introduction.

Virtual high-throughput screening (VHTS)[1-8] powered by density functional theory (DFT) coupled with machine learning (ML)[9-15] has shown promise to accelerate the discovery of materials. This acceleration is necessary because exploring large spaces of candidate materials introduces combinatorial challenges. Exemplary of the challenges that arise from combinatorial explosion is single-site inorganic catalyst design, where metals, ligands, and substrates all must be considered.[16-18] To address this challenge, ML has been applied to predict thermodynamic quantities to rapidly screen this combinatorial design space in both homogeneous[5, 19-23] and heterogeneous catalyst[24-30] design. Combined with active learning[31-33] and global optimization algorithms[34-36], catalysts with optimal catalytic properties can be quickly identified under a given mechanism. However, it is often difficult to experimentally characterize all mechanistically relevant intermediates due to their transient nature[37] and thus computational efforts towards reaction mechanism exploration are desired.[38-39] In this case, ML combined with automated VHTS workflows can accelerate the exploration of potential reactive intermediates and reaction pathways.[40-45]

Many promising catalysts are comprised of mid-row transition metals, which give rise to favorable reactivity due to their unpaired electrons and superior tunability in response to changing coordination environment.[46-53] However, these exact same characteristics of transition metal catalysts often lead to failed geometry optimizations due to converging to unintended geometries or unexpected electronic states.[54-55] Because we require the knowledge of all relevant intermediates to compute the full thermodynamic landscape of a catalyst or to explore multiple possible reaction mechanisms, the VHTS of catalysts is usually accompanied by high overall failure rates and wasted computational resources.

Recently, ML models have been developed to predict the computational cost[56-57] or suggest the most inexpensive density functional that still will be of reasonable accuracy[58] for a calculation to optimize the use of finite computational resources. These approaches, however, do not overcome wasted computational time due to their assumption that the calculations will eventually succeed. On the other hand, one can directly predict the likelihood of success for a calculation to avoid failed calculations. In our previous work[54], we built ML models to directly classify outcomes (i.e., success or failure) of transition metal complex geometry optimizations using the 2D molecular graph-based descriptors (i.e., revised-autocorrelations or RACs[59]) as inputs. While achieving 88% accuracy on the set-aside test data, the RACs-based models failed to generalize to chemical spaces that are distinct to the training data due to its explicit dependence on chemical compositions as inputs.

To overcome this issue, we introduced a dynamic classifier[54, 60], which monitors a geometry optimization on the fly and terminates a calculation if it is predicted to be unproductive. This neural network dynamic classifier takes step-series inputs of both the geometric and electronic structure features (e.g., energy gradient and Mulliken bond orders) and predicts the likelihood of success during a geometry optimization (Figure 1). Since this model uses incremental information gathered over the course of a DFT optimization, the dynamic classifier can generalize well across different chemical spaces.[54] This good transferability is particularly important in catalyst design, since we would like to only train a single model that works well for all reactive intermediates possibly involved in a reaction. A similar idea has also been recently adopted in predicting trajectories of molecular dynamics simulations[61-62] and the dynamic control of tokamak reactors[63-64].

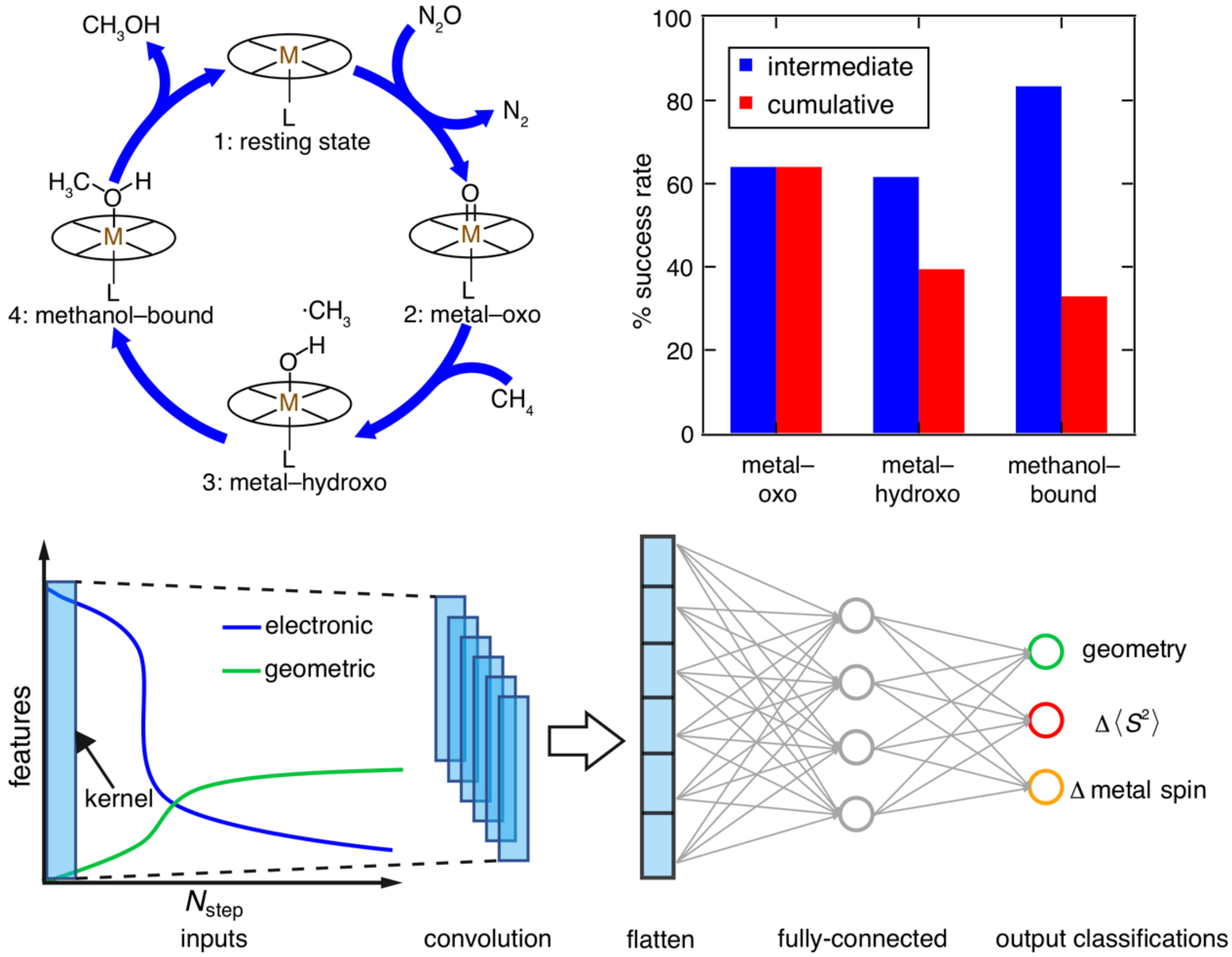

**Figure 1.** (top left) The radical rebound mechanism for direct partial oxidation of methane to methanol. The cycle proceeds clockwise from the resting state (**1**) to the metal−oxo intermediate (**2**) formed by two-electron oxidation with $N_2O$, followed by hydrogen atom transfer to form a metal–hydroxo intermediate (**3**), and rebound to form a methanol–bound intermediate (**4**). (top right) Success rate for geometry optimizations for each intermediate (blue) and cumulative success rate when a catalyst proceeds to each intermediate after success of the previous intermediate in the catalytic cycle (red). (bottom) Schematic of a multi-task dynamic classifier. Electronic structure and geometric features are collected from DFT geometry optimization and used as inputs to an ANN classifier with a convolutional layer followed by a fully connected layer. The model has multiple outputs that predicts calculation success with respect to geometry, $\langle S^2 \rangle$ deviation, and metal spin deviation.

In this work, we exploit and demonstrate the general nature of the transferability of our dynamic classifier for catalyst design. We show that a dynamic classifier can perform equivalently well on all reactive intermediates in a representative reaction. Despite being trained on only one reactive intermediate in a reaction cycle, this dynamic classifier generalizes to unseen intermediates within that same cycle. In addition, this dynamic classifier makes accurate predictions on reactive intermediates with distinct chemistry that are absent in the training data. We further incorporate uncertainty quantification when using dynamic classifier for job control, saving more than half of the computational resources that would have been wasted on failed calculations.

## 2. Results and discussion.

### 2.1 Generalizing the dynamic classifier across a catalytic cycle

The design of selective and active C−H activation catalysts for direct methane-to-methanol conversion remains a grand challenge.[65-66] Here, we focus on the radical rebound mechanism on representative Mn and Fe catalysts with macrocyclic tetradentate ligands, which have shown promise for exhibiting favorable thermodynamics for partial methane oxidation.[67-69] The "whole cycle" (*WC*) data set consists of a number of intermediates bound to these catalysts (Figure 1 and see Computational Details).[35] Starting from a resting state structure (**1**), a metal–oxo intermediate (**2**) is formed *via* two-electron oxidation with a terminal oxidant (here, $N_2O$). The newly formed terminal oxo can undergo hydrogen atom transfer step where a hydrogen atom is abstracted from $CH_4$ to form a metal‑hydroxo intermediate (**3**). Lastly, $CH_3\cdot$ recombines with the metal‑hydroxo intermediate to form a methanol‑bound intermediate (**4**). Thus, properties of both the resting state (**1**) and each of the three reactive intermediates (i.e., **2**, **3**, and **4**) must be obtained to evaluate the full thermodynamic landscape of a catalyst (see Computational Details). We focus here on only the reactive intermediates because we follow the convention of prior work[68] to evaluate (**1**) as a single point energy (i.e., without geometry optimization) on intermediate **2** with the oxo removed. Even if the success rate is high on each individual intermediate, the cumulative success rate can decay rapidly since multiple intermediates are necessary to obtain reaction energetics. We observe an overall success rate of only 33% for the whole catalytic cycle although all three intermediates have individual success rates ranging from 60% to 83% (Figure 1 and ESI Table S1).

We first train our dynamic classifier as a multi-task classification model for predicting three optimization outcomes, geometry, $\langle S^2 \rangle$ deviation, and metal spin deviation, on all three reactive intermediates in the *WC* set (see Computational Details). The first property, good geometry, corresponds to whether a structure optimizes to the intended connectivity expected for the structure and is general even to closed shell systems such as organic molecules. The latter two properties, $\langle S^2 \rangle$ deviation, and metal spin deviation, correspond to whether the structure has a large degree of spin contamination (i.e., $\langle S^2 \rangle$ differs from its expected value) or if the spin does not reside on the metal. Importantly, a geometry can be good while the spin is not localized to the metal or $\langle S^2 \rangle$ deviation is too large and vice versa. We focus on these three properties because they are the most common sources of failure or indicate low reliability of single-reference DFT results in VHTS.

The dynamic classifier systematically improves when the inputs consist of an increasing number of optimization steps (Figure 2 and ESI Figure S1). The model performs equivalently well on all three intermediates (ESI Figure S1). For the first few steps of the geometry optimization, the relatively poor model accuracy (ca. 0.85) on the geometry classification task may result in false negative predictions that incorrectly terminate calculations that could be expected to converge successfully. To overcome this challenge, we previously introduced a classification-model-specific uncertainty quantification metric, the latent space entropy (LSE)[54], to ensure high model confidence during prediction. Using the LSE as a guide for model uncertainty, we only act on model predictions if the LSE value is below a user-defined cutoff (here, 0.3). Using this requirement for model certainty, we achieve uniformly high model accuracy (i.e., > 0.95) for all optimization step numbers and intermediates for all three tasks (Figure 2 and ESI Figure S2). As the dynamic classifier is provided more information about the optimization (i.e., with an increasing number of steps), model confidence grows and the data fraction that falls below the LSE cutoff increases (Figure 2 and ESI Figure S2).[54]

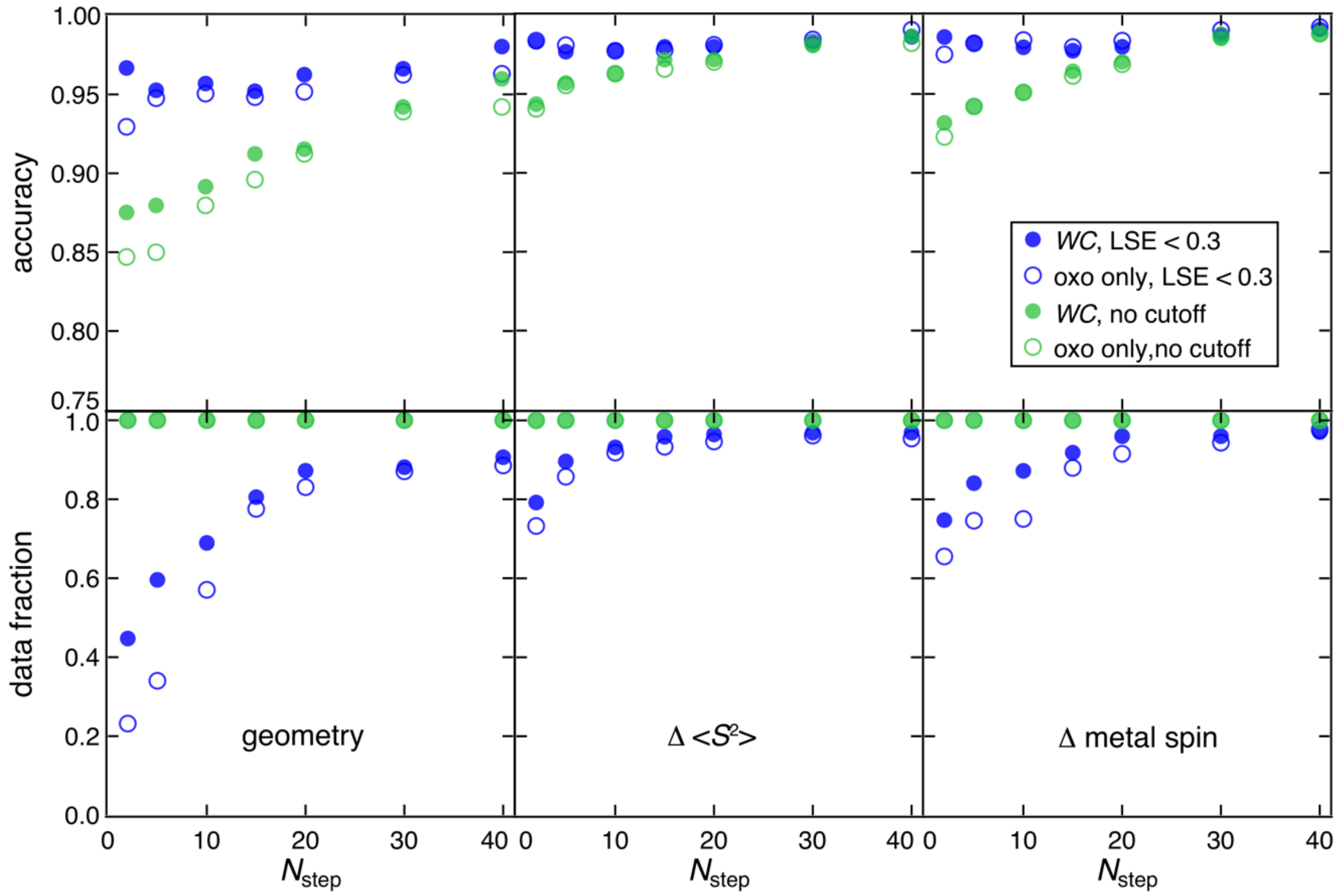

**Figure 2.** Model accuracy (top) and data fraction (bottom) versus the number of geometry optimization steps for the dynamic classifier at each $N_{\mathrm{step}}$ (i.e., 2, 5, 10, 15, 20, 30, and 40) evaluated on the set-aside test set of the *WC* set. The performance of each task: geometry (left), $\langle S^2 \rangle$ deviation (middle), and metal spin deviation (right), is reported separately. We report two sets of dynamic classifiers: one that is trained on all three intermediates in the *WC* set (solid circles), the other trained only on the metal–oxo intermediate in the *WC* set (open circles). Model performance is shown for both no model uncertainty control (green) and when we impose a LSE cutoff of 0.3 (blue).

We visualize the latent space (i.e., the outputs of the last hidden layer in a dynamic classifier) to understand why our dynamic classifier performs equivalently well on all three classification tasks simultaneously. We find that calculations corresponding to each failure mode cluster into distinct regions of the latent space (Figure 3). The bad calculations reside in a distinct portion of latent space from those that correspond to good calculations. The relative orientation of each of the three types of clusters is also intuitive. There is a cluster of calculations with both high $\langle S^2 \rangle$ deviation and metal spin deviation since these two failure modes both stem from the unexpected electronic structure and are typically concurrent (ESI Figure S3). Similarly, the boundary between calculations with good or bad geometry is the least well separated, consistent with this being the most challenging task among the three for our models to predict (Figure 2).

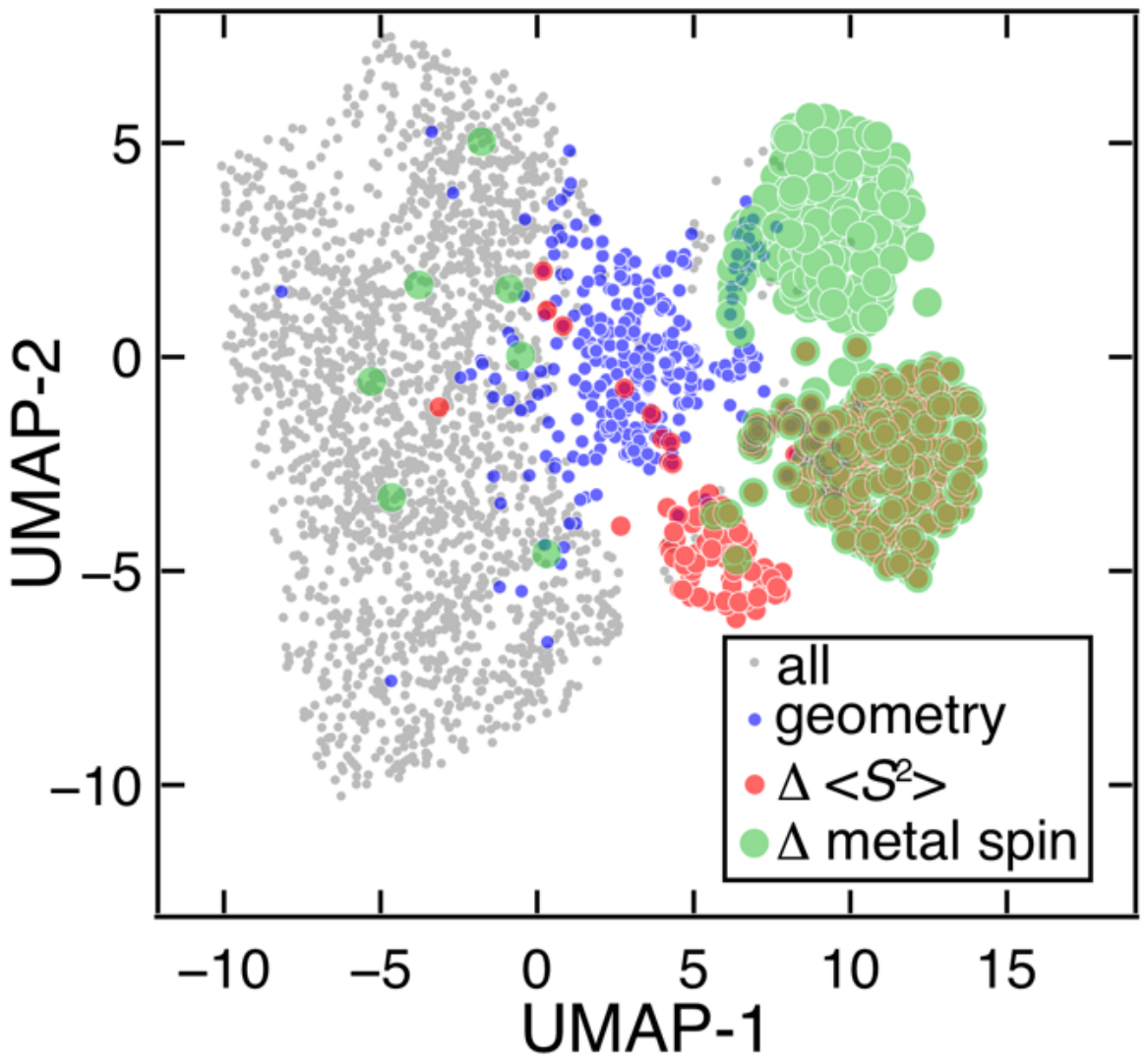

**Figure 3.** Uniform manifold approximation and projection[70] (UMAP) 2D visualization of the latent space of the multi-task dynamic classifier trained on 40 steps of geometry optimization trajectories of all three intermediates in the *WC* set. All data points from the *WC* set are shown in gray. Geometry optimizations that are labeled as bad are colored separately for each failure mode: red for geometry, blue for $\langle S^2 \rangle$ deviation, and green for metal spin deviation. Different sizes of circles are used only for the clear visualization of points around overlapping points.

Encouraged by the good performance of the dynamic classifier and good transferability offered by electronic structure inputs, we tested the dynamic classifier in a use case representative of a lower data regime. Here, we only train the dynamic classifier using the geometry optimization results obtained for the metal–oxo intermediate in the *WC* set (see

Computational Details). This oxo-only dynamic classifier performs comparably to the dynamic classifier trained on all three intermediates of the *WC* set (Figure 2). This good performance is observed despite the model having roughly 1/3 of the training data in the *WC* set and only learning from one intermediate out of the three. Specifically, the oxo-only dynamic classifier is within a margin of 1% for $\langle S^2 \rangle$ deviation and metal spin deviation classifications and within 3% for the geometry classification (Figure 4). The two sets of dynamic classifier models give nearly identical predictions on each individual set-aside test point in the *WC* set and have comparable latent space structures, which suggest that the dynamic classifier can learn similar information from only one intermediate in the reaction cycle (ESI Figures S4–S5). This observation suggests a promising reduction (i.e., to $1/N_{\text{intermediate}}$) in the number of necessary training data points for the dynamic classifier since multiple catalytic intermediates of a reaction must typically be screened.

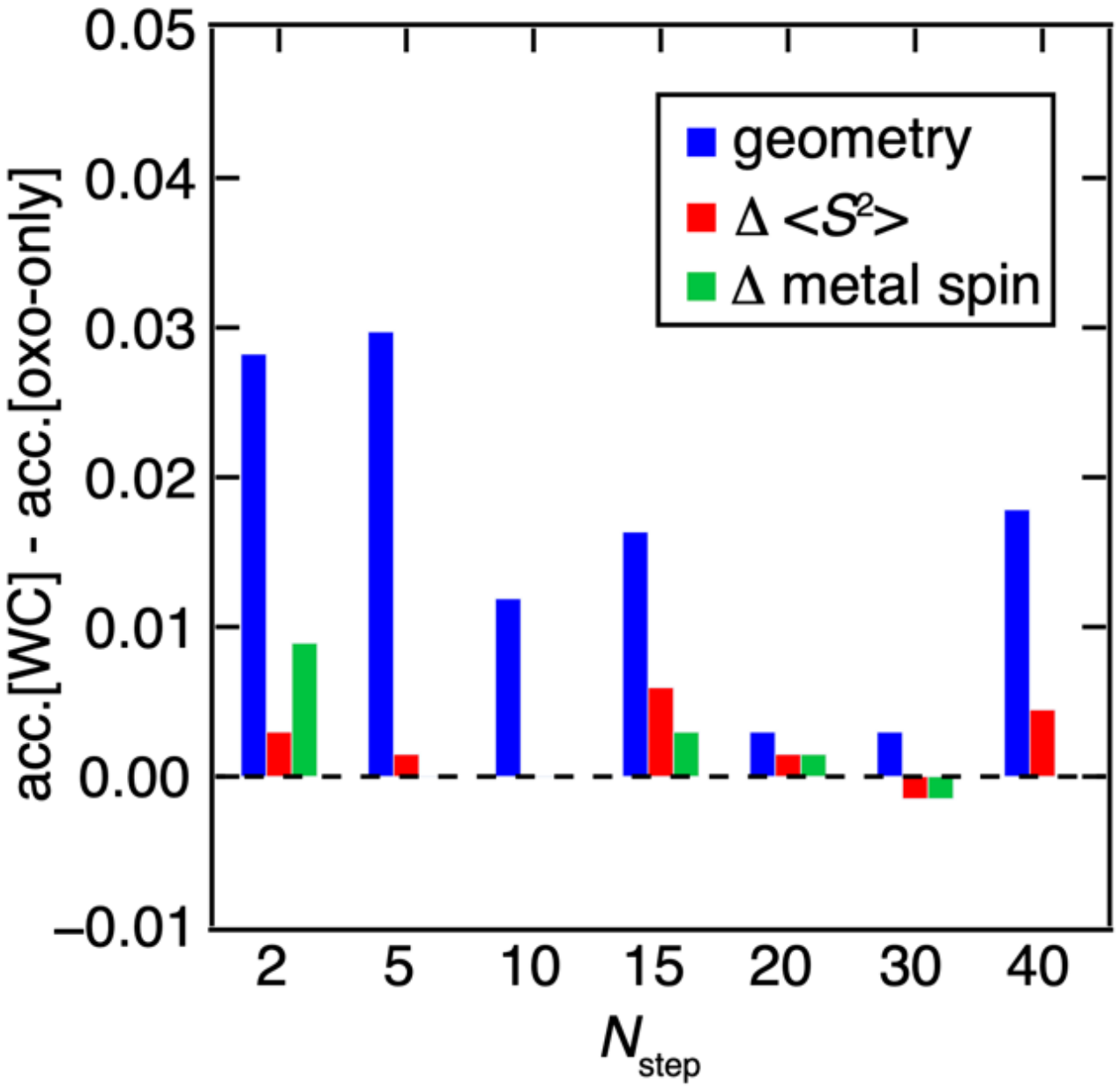

**Figure 4.** Difference in model accuracy (acc.) between the dynamic classifier trained on all three intermediates and the one trained on only the metal–oxo intermediate in the *WC* set at various optimization steps. The difference for all three tasks is shown separately: blue for geometry, red for $\langle S^2 \rangle$ deviation, and green for metal spin deviation. A dashed line is shown for no difference.

## 2.2 Transferability of the dynamic classification model to *out-of-distribution* catalysts

We next tested the transferability of the dynamic classifier to intermediates and catalysts beyond the initial set, which is essential for its use in catalyst discovery applications. To do so, we curated three additionaly data sets that are chemically distinct from the *WC* set but are applicable to catalyst screening focused on direct methane-to-methanol conversion. The first set is the functionalized whole cycle (*FWC*) data set, where tetradentate macrocycles were functionalized with electron-withdrawing and electron-donating groups to introduce Hammett tuning effects on catalyst energetics (Figure 5 and ESI Figure S6). In the *FWC* set, we functionalize all three reactive intermediates as in the *WC* set. In the Ru–oxo (*RO*) data set, we randomly sampled 300 metal–oxo species (metal = Mn, Fe) from the *WC* set and substituted the metal centers with Ru prior to re-optimization (Figure 5). We use the *RO* set as an example of catalyst design with isovalent metals, motivated by the fact that Ru compounds are promising catalysts for C–H bond activation and oxidation reactions.[71-72] Lastly, we introduce the carbonyl species (*CS*) data set, where we replace the oxo with a carbonyl ligand on all converged catalysts in the *WC* set (Figure 5). The *CS* set thus contains a representative off-cycle intermediate that could be generated in conditions of methane overoxidation and would be likely to poison the catalyst. Importantly, the metal-coordinating element (i.e., C) is distinct in the *CS* set from the other three sets. Since the chemical compositions of the four data sets (i.e., *WC*, *FWC*, *RO*, and *CS*) are distinct (i.e., either due to metal or coordinating species), we observe significantly different distributions for both their chemical-composition-based representation (e.g. RACs[59]) and electronic structure-based descriptors (e.g., Mulliken charges and bond orders, ESI Figures S7–S8).

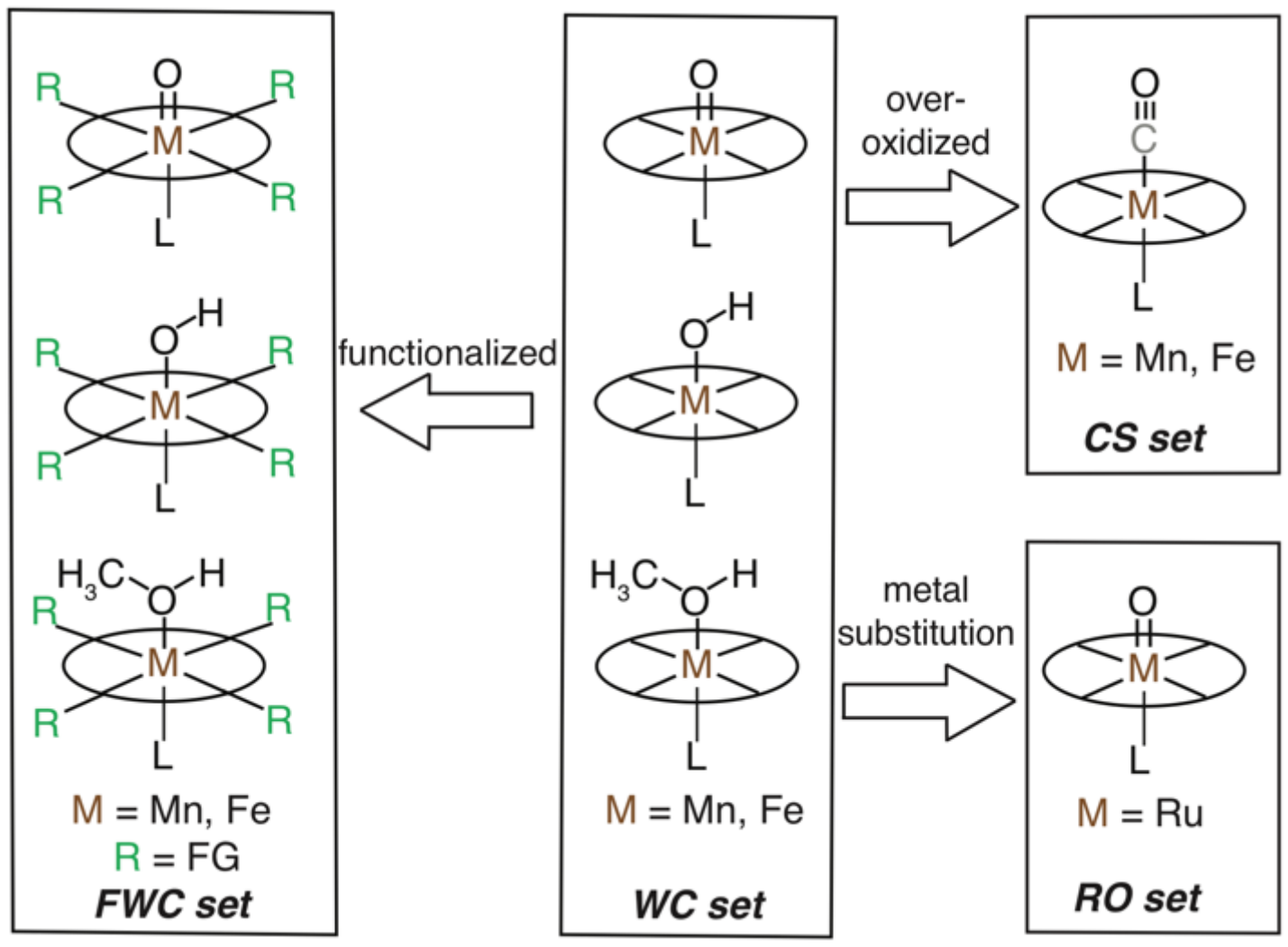

**Figure 5.** Schematic of *out-of-distribution* test data sets: The *FWC* set (left) is constructed by adding functional group (FG) on the rings and bridges of the base macrocycles in the *WC* set (middle). The *CS* set (top right) is constructed by changing the substrate on the metal to carbonyl, a common product when methane is over-oxidized. The *RO* set (bottom right) is constructed by substituting the metal (i.e., Mn or Fe) in the *WC* set with Ru.

We find that the dynamic classifier trained only on the metal–oxo intermediate in the *WC* set shows exceptional transferability to all three test sets despite differences in chemical composition. The accuracy for the most difficult prediction task (i.e., geometry classification) is comparable among all four data sets (i.e., *WC*, *FWC*, *RO*, and *CS*). Namely, we observe geometry classification performance accuracy to be within 5% for all four data sets, regardless of the number of steps used for dynamic classification (Figure 6). In addition, the accuracy for the other two tasks, $\langle S^2 \rangle$ deviation and metal spin deviation, is identical for the *out-of-distribution FWC, RO, CS* sets relative to the *in-distribution WC* set (ESI Figure S9). More interestingly, the model confidence (i.e., average LSE) of the dynamic classifier on the three *out-of-distribution* test sets is comparable relative to the *in-distribution WC* set-aside test set for all three classification tasks (Figure 6 and ESI Figure S9). This observation suggests that our oxo-only dynamic classifier has similar confidence for making predictions on a compound that is chemically distinct from the training data due to good transferability across chemical compositions.

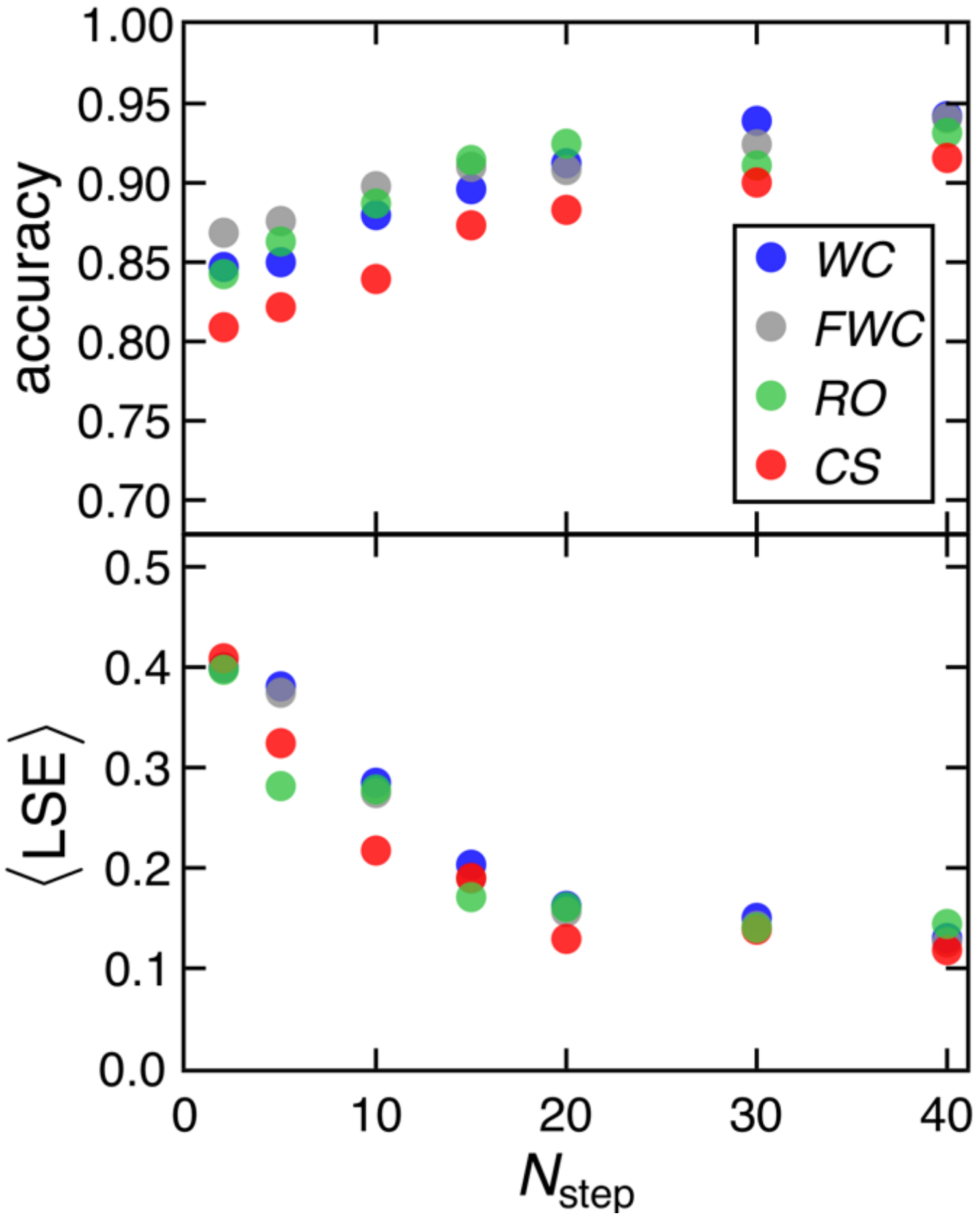

**Figure 6.** Accuracy (top) and the average LSE (bottom) for the geometry classification task for the set-aside test set in *WC* set (blue) and three *out-of-distribution* test sets (*FWC* in gray, *RO* in green, and *CS* in red) with increasing number of optimization steps, $N_{\text{step}}$. The dynamic classifier was trained only on the metal–oxo intermediate in the *WC* set.

We can further rationalize the transferability of our dynamic classifier from an analysis of the inputs to the model. These inputs are electronic structure and geometric features generated from DFT calculations on the fly, which make them agnostic to catalyst chemical composition. As a result, all *out-of-distribution* intermediates, despite being chemically distinct from the training complexes, reside within the 2D projected convex hull spanned by the metal–oxo intermediate of the *WC* set in the latent space of the dynamic classifier (Figure 7). Therefore, for a new geometry optimization trajectory, the dynamic classifier can be expected to have good training data support even if the specific intermediate or catalyst has not been seen by the model. This is a consequence of our use of electronic structure and geometric features and would not have been possible with a chemical-composition-based representation (e.g., RACs). With chemical-composition-based representations, the *out-of-distribution* intermediates reside in different regions of latent space, extending beyond the 2D convex hull spanned by the metal–oxo intermediate of the *WC* set (ESI Figure S6).

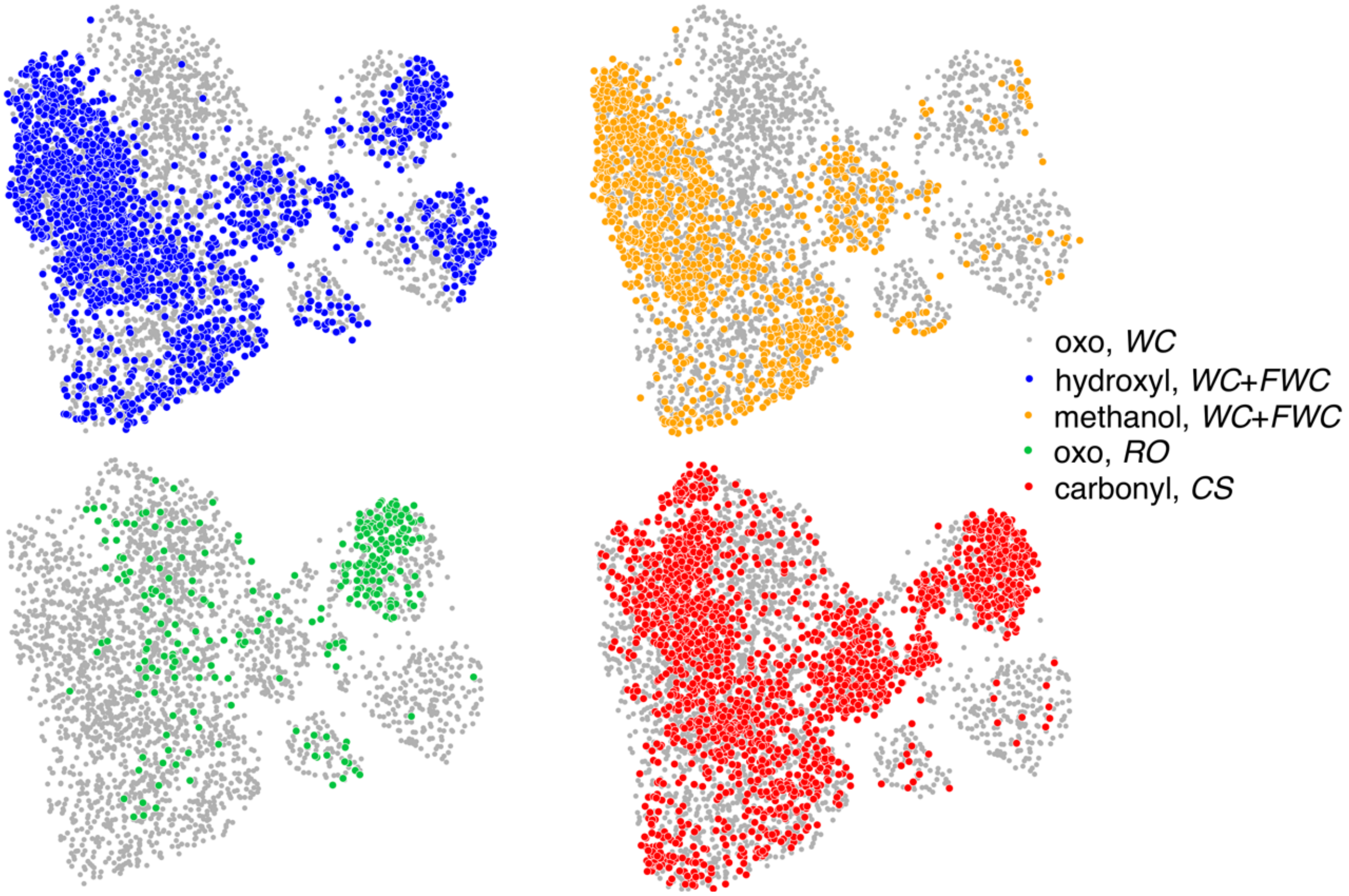

**Figure 7.** UMAP 2D visualization of the latent space for different intermediates of the multi-task dynamic classifier trained on 40 steps of geometry optimization trajectories of the metal–oxo intermediate in the *WC* set (gray). Multiple intermediates in different data sets are shown separately: metal–hydroxo intermediate in the *WC* and *FWC* set (blue, top left), metal–methanol intermediate in the *WC* and *FWC* set (orange, top right), Ru–oxo intermediate in the *RO* set (green, bottom left), and metal–carbonyl intermediate in the *CS* set (red, bottom right).

Another reason for the good transferability is expected to be due to the fact that the dynamic classifier learns from trends in how electronic structure and geometric features evolve over the course of a geometry optimization rather than solely from the value of each feature at a single optimization step. This is inherent to the dynamic classifier model architecture, which involves a 1D convolution through the dimension of the optimization step. For example, an IS Mn(II)–oxo complex in the *WC* set and a HS Mn(II)–methanol complex in the *FWC* set have similar trends in their electronic structure trajectories (e.g., for Mulliken bond valence on the metal) but distinct values of this property. However, the dynamic classifier can correctly classify both optimization trajectories as good with high confidence (LSE < 0.1) even though their metal bond valences lie at opposite extrema of the distribution (ESI Figure S7).

For a convolutional neural network, one can visualize the model focus while making predictions using gradient class activation map (GCAM)[73]. Here, the GCAM focus on the two trajectories are comparable, indicating that the dynamic classifier makes the same prediction for the same reason. In contrast, an IS Fe(III)–carbonyl complex in the *CS* set has distinct trends in its trajectory despite similar values of the metal bond valence to the IS Mn(II)–methanol complex. For this HS Fe(III)–carbonyl complex, however, the distal axial ligand dissociates and produces a bad geometry. This time, the dynamic classifier confidently (LSE < 0.1) predicts this Fe(III)–carbonyl compound to result in a bad geometry by recognizing distinct fluctuations in properties during geometry optimization. Interestingly, GCAM reveals that the dynamic classifier primarily focuses on the later portion of the trajectory (i.e., step > 18), which corresponds to the second peak and decay in the metal bond valence trajectory data. This point is approximately at the point in the optimization where dissociation of the distal axial ligand starts to occur.

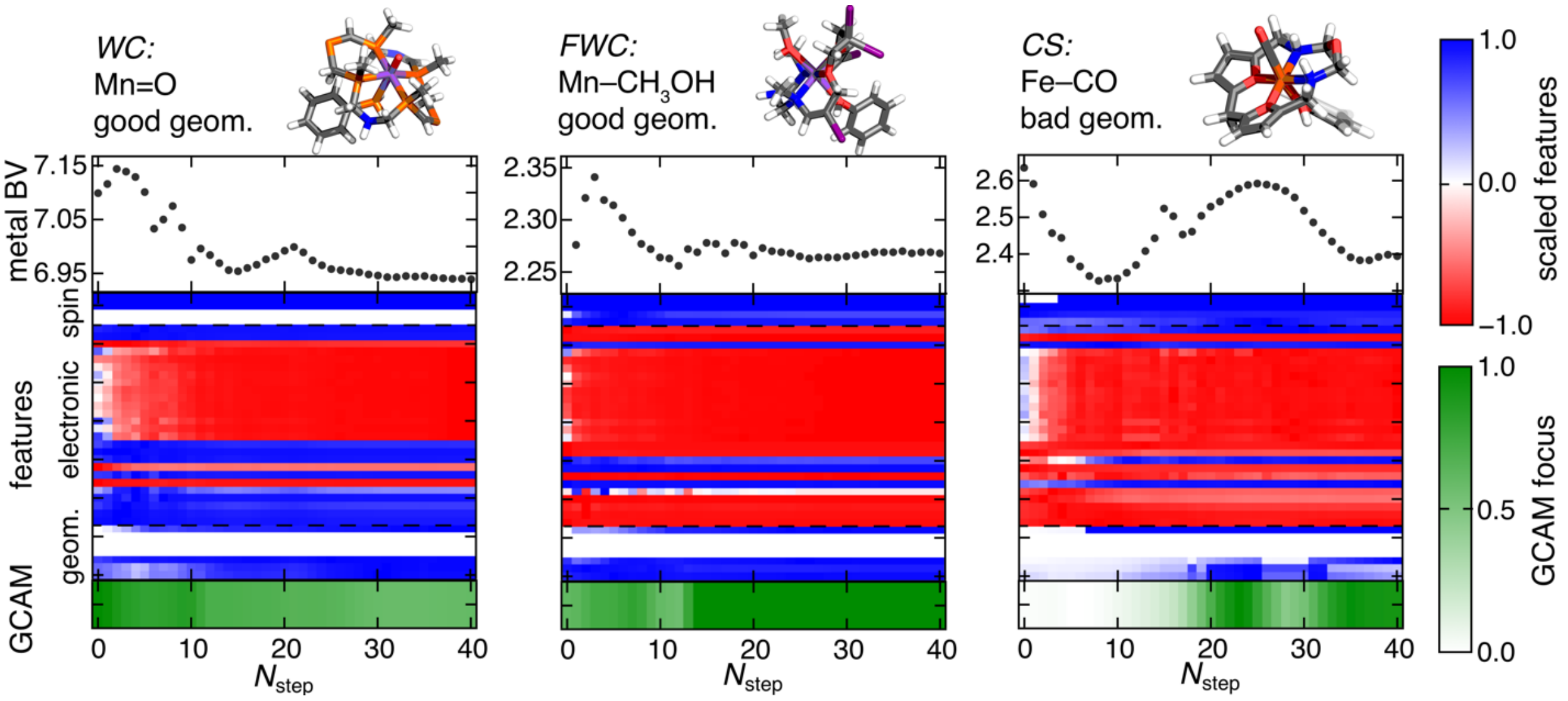

**Figure 8.** Metal bond valence (BV, top), scaled dynamic features (middle), and GCAM focus of the geometry classification task versus the number of steps of optimizations. The scaled dynamic features and GCAM focus are colored following the color bars (right). Properties are shown for three example complexes: a Mn–oxo complex in the *WC* set with a final good geometry (left), a functionalized Mn–methanol complex in the *FWC* set with a final good geometry (middle), and a Fe–carbonyl complex in the *CS* set with a final bad geometry (axial ligand dissociated, right). The dynamic classifier used for GCAM analysis was trained on 40 steps of geometry optimization trajectories obtained only from the metal–oxo intermediate in the *WC* set. In all three cases, the dynamic classifier makes the correct prediction with high confidence (LSE < 0.1).

After introducing an LSE cutoff of 0.3 as a UQ metric, we achieve uniformly high accuracy for all prediction tasks, i.e., > 0.90 for geometry and > 0.97 for both $\langle S^2 \rangle$ and metal spin deviation, for all four data sets at all optimization steps (ESI Figure S10). This consistent performance is surprising because the dynamic classifier was only trained on the metal–oxo intermediates in the *WC* set. Overall, this high accuracy leads to a reduction of more than 1/2 of the computational time that would have been wasted due to failed calculations along with a negligible false negative rate (< 2%) for each of the four data sets (Figure 9). Thanks to the uniformly good performance across chemically distinct *out-of-distribution* test sets especially when paired with uncertainty quantification, the dynamic classifier can be expected to be transferable for other mechanistic studies and catalyst screening efforts. We anticipate the dynamic classifier to be readily transferable across catalysts with different metal, oxidation state, spin state, and ligand environment but would expect it to require additional training data when it is applied to catalysts with different coordination number and geometry type.

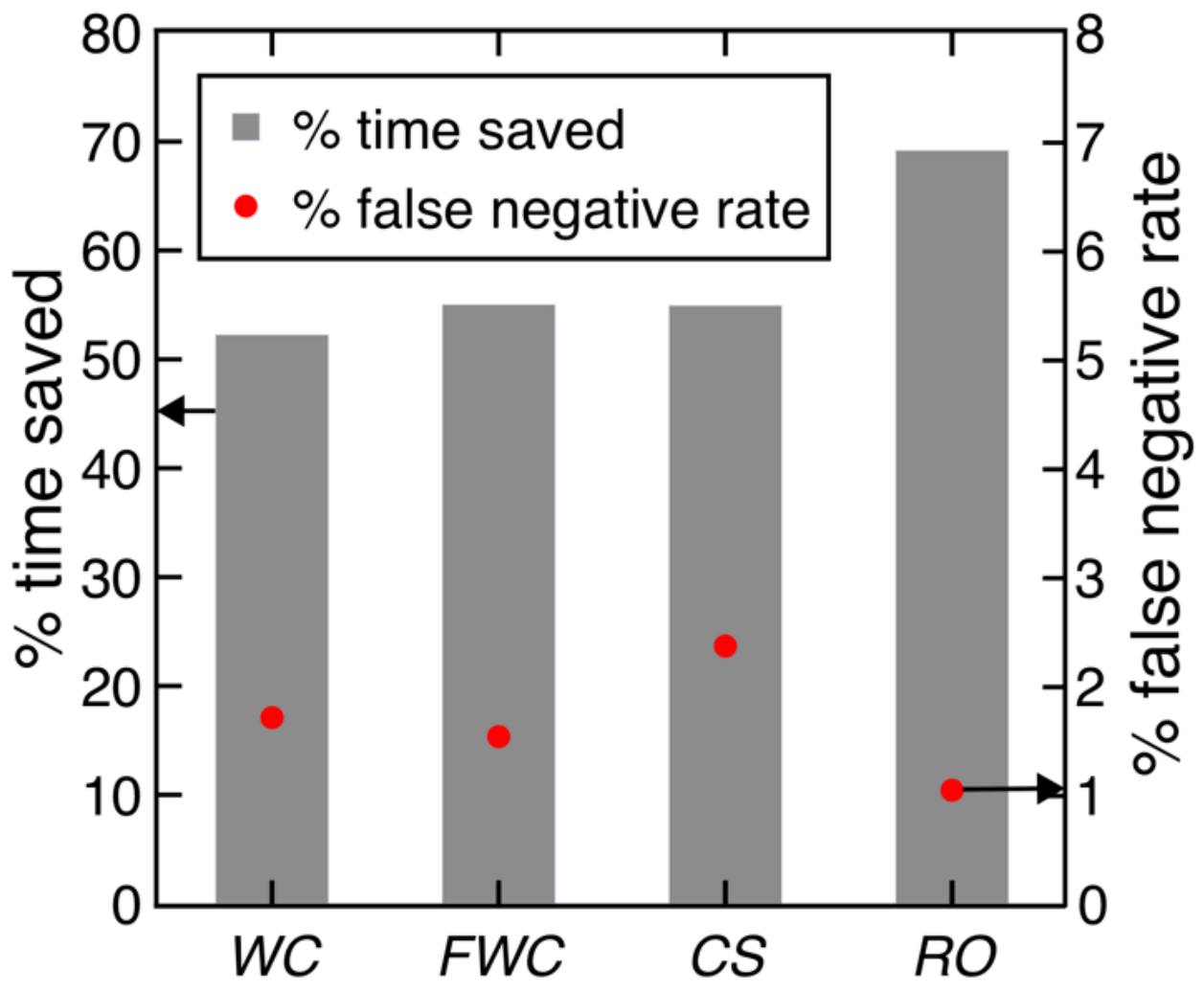

**Figure 9.** Percentage of time saved from bad calculations (gray bars, y-axis on the left) and false negative rate (red circles, y-axis on the right) for the uncertainty-aware dynamic job control. Results are reported for the set-aside test set in *WC* set and three out-of-distribution test sets (i.e., *FWC*, *RO*, and *CS*). The dynamic classifier was trained on 40 steps of geometry optimization trajectories obtained on the metal–oxo intermediate in the *WC* set. An LSE cutoff of 0.3 was used to make model predictions.

## 3. Conclusions

Computational catalysis requires the rapid screening of both catalyst compositions and intermediates. As these screening efforts are increasingly carried out with automated workflows, it becomes essential to anticipate and detect when calculations fail. To address this, we built a dynamic classifier to predict geometry optimization outcomes on the fly for reactive intermediates. We demonstrated our approach on the challenging reaction of direct methane-to-methanol conversion via a radical-rebound mechanism. We demonstrate that the dynamic classifier trained on all reactive intermediates exhibits good, balanced performance on each intermediate. Encouraged by the model's good transferability across intermediates, we tested the model in a lower data regime where only the metal–oxo intermediate was included in the training data. This oxo-only dynamic classifier performed similarly well compared to the original model. This observation is general, suggesting the promise of reducing the required training data to $1/N_{\text{intermediate}}$ in practical applications to complex reaction networks. A proposed workflow motivated by this observation is to only train the dynamic classifier on the first reactive intermediate of a reaction cycle and then apply that model for all additional reactive intermediates to accelerate screening of the full catalytic cycle.

In addition to expected catalytic intermediates, a true test of transferability for computational catalysis requires that the model generalize to reactive intermediates with distinct chemistry. We evaluated model performance on catalysts that were functionalized with small functional groups frequently employed in Hammett tuning, those with substituted transition metals (i.e., Ru instead of Fe), and intermediates with distinct metal-coordinating atoms (here, metal–carbonyl). For all three sets, we demonstrated that the oxo-only dynamic classifier generalized well on these intermediates with distinct chemical composition. We rationalized the transferability of dynamic classifier in two ways. First, the dynamic classifier only uses electronic structure and geometric features generated during DFT geometry optimizations. Thus, the model can be expected to be transferable because DFT-based descriptors are likely to be more comparable than chemical-composition-based ones. Second, the dynamic classifier utilizes a convolution layer for step-series features generated during an optimization, making it focus on the trends of a trajectory rather than the value of each feature that we observed to differ more significantly between catalysts.

We incorporated an uncertainty quantification metric in the form of the latent space entropy to ensure that the oxo-only dynamic classifier made predictions only where it was most confident. Using this approach, we demonstrated a greater than 50% reduction in computational time to carry out catalyst screening by avoiding unsuccessful calculations along with negligible false negative predictions (< 2%) for all intermediates considered in this work. The uniformly high reduction rate with few false negatives highlights how this dynamic classifier model is ready for use to improve the robustness of automated workflows, perhaps even beyond that which can normally be achieved via manual intervention. This uncertainty-aware dynamic classifier represents a promising approach to both accelerate and improve the fidelity of VHTS, and we expect our approach to be general to an even wider range of materials and catalyst screening studies than the mid-row transition metal catalysts studied here.

## 4. Computational details

### 4.1 DFT geometry optimizations

Gas-phase geometry optimizations and single-point energy calculations were performed using density functional theory (DFT) with a development version of TeraChem v1.9.[74-75] The B3LYP[76-78] global hybrid functional with the empirical D3 dispersion correction[79] using Becke−Johnson damping[80] was employed for all calculations. The LACVP* composite basis set was employed throughout this work, which consists of a LANL2DZ effective core potential[81-82] for Mn, Fe, Ru, Br, and I and the 6-31G* basis[83] for all other atoms. All singlet calculations were carried out in a spin-restricted formalism whereas all other spin states were performed in the unrestricted formalism. As motivated by prior work[68], we simulated the metal–hydroxo intermediate by majority-spin radical addition to the metal–oxo intermediate. For other reactive intermediates (e.g. resting state and methanol-bound intermediates), we conserved the metal–oxo spin state. Level shifting[84] of 0.25 Ha was applied to both majority and minority spin virtual orbitals to aid self-consistent field (SCF) convergence. Geometry optimizations were carried out with the translation rotation internal coordinate (TRIC) optimizer[85] using the L-BFGS algorithm with default convergence thresholds of maximum energy gradient of $4.5\times10^{-4}$ hartree/bohr and energy difference between steps of $10^{-6}$ hartree.

Job submission was automated by molSimplify[86-87] with a 24 h wall time limit per run with up to five resubmissions. Geometry optimizations were carried out with geometry checks[54] prior to each resubmission, and structures that failed any check were labeled as failed calculations (ESI Table S2). Open-shell calculations were also deemed failed calculations in the data set following established protocols[19, 54-55]: i) if the expectation value of the $S^2$ operator deviated from its expected value of $S(S + 1)$ by >1 $\mu_B^2$ or ii) the combined Mulliken spin density on the metal and oxygen differed from the spin multiplicity by >1 $\mu_B$ (ESI Table S3).

### 4.2 Data sets

We calculated the radical rebound mechanism[88] for methane-to-methanol conversion on mononuclear Mn and Fe catalysts with realistic tetradentate macrocycles constructed from known ligand fragments[35] (ESI Figure S11). For these catalysts, two resting state oxidation states, M(II) and M(III), in their corresponding spin states were considered (ESI Table S4). For this catalytic cycle, we optimize three catalytic intermediates: metal–oxo, metal–hydroxo, and the methanol-bound species. The initial geometries for metal−oxo species were constructed using molSimplify,[86] which uses OpenBabel[89-90] as a backend to interpret SMILES strings for constructed tetradentate macrocycles. All metal−hydroxo geometries were generated by adding an H atom to the optimized metal−oxo structure, and all methanol-bound intermediates were generated by adding a methyl group to the optimized metal−hydroxo structures using a custom script in molSimplify, as in prior work[68] (ESI Figure S12). The workflow starts by optimizing the metal−oxo geometry, and if this structure or a subsequent intermediate fails, downstream intermediate optimizations are not attempted (ESI Figure S13). We refer to this combined data set of metal−oxo, metal−hydroxo, and methanol-bound intermediates as the "whole cycle" (*WC*) data set (ESI Table S1).

Starting from the *WC* data set, we generated three data sets inspired by common strategies and potential difficulties (i.e., overoxidation) in catalyst discovery: 1) the functionalized whole cycle (*FWC*) data set, where tetradentate macrocycles were functionalized with electron withdrawing and electron donating groups to introduce Hammett tuning effects, 2) the Ru-oxo species (*RO*) data set, where the metal (Mn or Fe) of 300 randomly sampled metal-oxo species in *WC* is substituted by Ru, and 3) the carbonyl species (*CS*) data set, where a carbonyl ligand replaces any converged oxo moiety in catalysts from the *WC* set (ESI Table S5). For the *FWC*, *RO*, and *CS* data sets, we follow the same procedure for geometry optimizations used for the *WC* data set. Additionally, the metal−oxo, metal−hydroxo, and methanol-bound intermediates in the *FWC* set were computed following the same workflow as in *WC* set (ESI Figures S12–S13). Each Ru-oxo complex in *RO* was initialized by a direct substitution of Mn or Fe with Ru from the initial geometry of an metal–oxo intermediate generated with molSimplify, with an initial Ru=O bond length of 1.65 Å. Each metal–carbonyl complex in the *CS* set was initialized by a direct substitution of CO in place of the oxo moiety from the initial geometry of metal–oxo intermediate generated with molSimplify, with an initial metal–C bond length of 2.10 Å, C–O bond length of 1.13 Å, and metal–C–O angle of 180º.

**4.3 ML models and representations**

As in prior work[54, 91], we train convolutional neural network dynamic classifiers using step-series electronic structure and geometric information generated during DFT geometry optimizations to directly predict the final classification outcomes of geometry fitness, $\langle S^2 \rangle$, and spin density deviation (ESI Table S6 and Figure S14). The 28 electronic structure descriptors were computed from the Mulliken charge, bond order matrix, and the energy gradient of a complex (ESI Table S6). These properties were focused on components directly involved in the first-coordination sphere along with any long-range behavior captured by singular value decomposition of these quantities (ESI Table S6). The 7 geometric descriptors include the bond lengths and angular deviation from an ideal octahedron as well as the distortion of individual ligand (ESI Table S1). We trained two sets of multi-task dynamic classifiers: one on all three intermediates (i.e., metal−oxo, metal−hydroxo, and methanol-bound) of the *WC* data set, and the other only on the metal−oxo species subset of the *WC* data set. For all ML models, we adopted same sets of hyperparameters used in our prior work[54] and a random 80/20 train/test split, with 20% of the training data (i.e., 16% overall) used as a validation set (ESI Table S7). All ML models were trained using Keras[92] with Tensorflow[93] as a backend, using the Adam optimizer up to 2,000 epochs, and dropout, batch normalization, and early stopping to avoid over-fitting.

## ASSOCIATED CONTENT

**Supplementary Information**. Geometry metrics and electronic structure cutoffs for calculation failure; Performance of the dynamic classifier for each intermediate in the *WC* set at each prediction task; Venn diagrams and job statistics for calculation failure modes; Prediction differences for two sets of dynamic classifiers; UMAP visualization for oxo-only dynamic classifier for RACs-155; Functional groups used in the *FWC* set; Histograms of electronic structure and geometric features; Performance of oxo-only dynamic classifier on all four data sets; Strategies of constructing tetradentate macrocycles and oxidation state and spin state of the catalysts; Workflow of computing intermediates in a reaction cycle; Architecture and hyperparameters of the multi-task dynamic classifier; dynamic classifier model h5 files; electronic structure and geometric feature json files; optimized geometries for catalysts in xyz files; data csv files.

AUTHOR INFORMATION

**Corresponding Author**

*email:hjkulik@mit.edu

**Notes**

The authors declare no competing financial interest.

ACKNOWLEDGMENT

This work was primarily supported by the Office of Naval Research under grant number N00014-20-1-2150. The work was also supported by the National Science Foundation under grant number CBET-1846426 and a National Science Foundation Graduate Research Fellowship under Grant #1122374 (to A.N.). C.D. was partially supported by a seed fellowship from the Molecular Sciences Software Institute under NSF grant OAC-1547580. Workflow development was supported in part by the United States Department of Energy grant number DE-NA0003965 and DARPA grant number D18AP00039. Additional support was provided by an AAAS Marion Milligan Mason Award and an Alfred P. Sloan Fellowship in Chemistry. H.J.K. holds a Burroughs Wellcome Fund Career Award at the Scientific Interface, which supported this work.

The authors thank Adam H. Steeves for providing a critical reading of the manuscript.

REFERENCES

1. Shu, Y. N.; Levine, B. G., Simulated Evolution of Fluorophores for Light Emitting Diodes. *J. Chem. Phys.* **2015,** *142* (10), 104104.
2. Gomez-Bombarelli, R.; Aguilera-Iparraguirre, J.; Hirzel, T. D.; Duvenaud, D.; Maclaurin, D.; Blood-Forsythe, M. A.; Chae, H. S.; Einzinger, M.; Ha, D. G.; Wu, T.; Markopoulos, G.; Jeon, S.; Kang, H.; Miyazaki, H.; Numata, M.; Kim, S.; Huang, W. L.; Hong, S. I.; Baldo, M.; Adams, R. P.; Aspuru-Guzik, A., Design of Efficient Molecular Organic Light-Emitting Diodes by a High-Throughput Virtual Screening and Experimental Approach. *Nat. Mater.* **2016,** *15* (10), 1120-+.
3. Kanal, I. Y.; Owens, S. G.; Bechtel, J. S.; Hutchison, G. R., Efficient Computational Screening of Organic Polymer Photovoltaics. *J. Phys. Chem. Lett.* **2013,** *4* (10), 1613-1623.
4. Vogiatzis, K. D.; Polynski, M. V.; Kirkland, J. K.; Townsend, J.; Hashemi, A.; Liu, C.; Pidko, E. A., Computational Approach to Molecular Catalysis by 3d Transition Metals: Challenges and Opportunities. *Chem. Rev.* **2018,** *119*, 2453-2523.
5. Foscato, M.; Jensen, V. R., Automated in Silico Design of Homogeneous Catalysts. *ACS Catal.* **2020,** *10* (3), 2354-2377.
6. Curtarolo, S.; Hart, G. L.; Nardelli, M. B.; Mingo, N.; Sanvito, S.; Levy, O., The High-Throughput Highway to Computational Materials Design. *Nat. Mater.* **2013,** *12* (3), 191-201.
7. Ong, S. P.; Richards, W. D.; Jain, A.; Hautier, G.; Kocher, M.; Cholia, S.; Gunter, D.; Chevrier, V. L.; Persson, K. A.; Ceder, G., Python Materials Genomics (pymatgen): A Robust, Open-Source Python Library for Materials Analysis. *Comput. Mater. Sci.* **2013,** *68*, 314-319.
8. Nørskov, J. K.; Bligaard, T., The Catalyst Genome. *Angew. Chem., Int. Ed. Engl.* **2013,** *52* (3), 776-777.
9. Meredig, B.; Agrawal, A.; Kirklin, S.; Saal, J. E.; Doak, J.; Thompson, A.; Zhang, K.; Choudhary, A.; Wolverton, C., Combinatorial Screening for New Materials in Unconstrained Composition Space with Machine Learning. *Phys. Rev. B* **2014,** *89* (9), 094104.
10. Dral, P. O., Quantum Chemistry in the Age of Machine Learning. *J. Phys. Chem. Lett.* **2020,** *11* (6), 2336-2347.
11. Janet, J. P.; Duan, C.; Nandy, A.; Liu, F.; Kulik, H. J., Navigating Transition-Metal Chemical Space: Artificial Intelligence for First-Principles Design. *Acc. Chem. Res.* **2021,** *54* (3), 532-545.
12. Butler, K. T.; Davies, D. W.; Cartwright, H.; Isayev, O.; Walsh, A., Machine Learning for Molecular and Materials Science. *Nature* **2018,** *559* (7715), 547-555.
13. Chen, A.; Zhang, X.; Zhou, Z., Machine learning: Accelerating materials development for energy storage and conversion. *InfoMat* **2020,** *2* (3), 553-576.
14. Pollice, R.; Gomes, G. D.; Aldeghi, M.; Hickman, R. J.; Krenn, M.; Lavigne, C.; Lindner-D'Addario, M.; Nigam, A.; Ser, C. T.; Yao, Z. P.; Aspuru-Guzik, A., Data-Driven Strategies for Accelerated Materials Design. *Acc. Chem. Res.* **2021,** *54* (4), 849-860.
15. Ceriotti, M.; Clementi, C.; von Lilienfeld, O. A., Introduction: Machine Learning at the Atomic Scale. *Chem. Rev.* **2021,** *121* (16), 9719-9721.

16. Nandy, A.; Duan, C. R.; Taylor, M. G.; Liu, F.; Steeves, A. H.; Kulik, H. J., Computational Discovery of Transition-metal Complexes: From High-throughput Screening to Machine Learning. *Chem. Rev.* **2021,** *121* (16), 9927-10000.
17. Vogiatzis, K. D.; Polynski, M. V.; Kirkland, J. K.; Townsend, J.; Hashemi, A.; Liu, C.; Pidko, E. A., Computational Approach to Molecular Catalysis by 3d Transition Metals: Challenges and Opportunities. *Chem. Rev.* **2018,** *119* (4), 2453-2523.
18. Kitchin, J. R., Machine learning in catalysis. *Nat. Catal.* **2018,** *1* (4), 230-232.
19. Nandy, A.; Zhu, J.; Janet, J. P.; Duan, C.; Getman, R. B.; Kulik, H. J., Machine Learning Accelerates the Discovery of Design Rules and Exceptions in Stable Metal-Oxo Intermediate Formation. *ACS Catal.* **2019,** *9*, 8243-8255.
20. Cordova, M.; Wodrich, M. D.; Meyer, B.; Sawatlon, B.; Corminboeuf, C., Data-Driven Advancement of Homogeneous Nickel Catalyst Activity for Aryl Ether Cleavage. *ACS Catal.* **2020,** *10* (13), 7021-7031.
21. Kalikadien, A. V.; Pidko, E. A.; Sinha, V., ChemSpaX: exploration of chemical space by automated functionalization of molecular scaffold. *Digital Discovery* **2022,** *1* (1), 8-25.
22. Song, Z. L.; Zhou, H. Y.; Tian, H.; Wang, X. L.; Tao, P., Unraveling the energetic significance of chemical events in enzyme catalysis via machine-learning based regression approach. *Commun. Chem.* **2020,** *3* (1).
23. Gensch, T.; Gomes, G. D.; Friederich, P.; Peters, E.; Gaudin, T.; Pollice, R.; Jorner, K.; Nigam, A.; Lindner-D'Addario, M.; Sigman, M. S.; Aspuru-Guzik, A., A Comprehensive Discovery Platform for Organophosphorus Ligands for Catalysis. *J. Am. Chem. Soc.* **2022**.
24. Tran, K.; Ulissi, Z. W., Active Learning Across Intermetallics to Guide Discovery of Electrocatalysts for $CO_2$ Reduction and $H_2$ Evolution. *Nat. Catal.* **2018,** *1* (9), 696.
25. Ulissi, Z. W.; Medford, A. J.; Bligaard, T.; Nørskov, J. K., To Address Surface Reaction Network Complexity Using Scaling Relations Machine Learning and DFT Calculations. *Nat. Commun.* **2017,** *8*, 14621.
26. Pellizzeri, S.; Barona, M.; Bernales, V.; Miró, P.; Liao, P.; Gagliardi, L.; Snurr, R. Q.; Getman, R. B., Catalytic Descriptors and Electronic Properties of Single-Site Catalysts for Ethene Dimerization to 1-Butene. *Catal. Today* **2018,** *312*, 149-157.
27. Vogiatzis, K. D.; Haldoupis, E.; Xiao, D. J.; Long, J. R.; Siepmann, J. I.; Gagliardi, L., Accelerated Computational Analysis of Metal–Organic Frameworks for Oxidation Catalysis. *The Journal of Physical Chemistry C* **2016,** *120* (33), 18707-18712.
28. Rosen, A. S.; Notestein, J. M.; Snurr, R. Q., Structure–Activity Relationships That Identify Metal–Organic Framework Catalysts for Methane Activation. *ACS Catal.* **2019,** *9* (4), 3576-3587.
29. Meyer, B.; Sawatlon, B.; Heinen, S.; von Lilienfeld, O. A.; Corminboeuf, C., Machine Learning Meets Volcano Plots: Computational Discovery of Cross-Coupling Catalysts. *Chem. Sci.* **2018,** *9* (35), 7069-7077.
30. Xu, J. Y.; Cao, X. M.; Hu, P., Perspective on computational reaction prediction using machine learning methods in heterogeneous catalysis. *Phys. Chem. Chem. Phys.* **2021,** *23* (19), 11155-11179.
31. Zhong, M.; Tran, K.; Min, Y.; Wang, C.; Wang, Z.; Dinh, C.-T.; De Luna, P.; Yu, Z.; Rasouli, A. S.; Brodersen, P.; Sun, S.; Voznyy, O.; Tan, C.-S.; Askerka, M.; Che, F.; Liu, M.; Seifitokaldani, A.; Pang, Y.; Lo, S.-C.; Ip, A.; Ulissi, Z.; Sargent, E. H., Accelerated Discovery of $CO_2$ Electrocatalysts Using Active Machine Learning. *Nature* **2020,** *581* (7807), 178-183.

32. Yang, W. H.; Fidelis, T. T.; Sun, W. H., Machine Learning in Catalysis, From Proposal to Practicing. *ACS Omega* **2020,** *5* (1), 83-88.
33. Flores, R. A.; Paolucci, C.; Winther, K. T.; Jain, A.; Torres, J. A. G.; Aykol, M.; Montoya, J.; Norskov, J. K.; Bajdich, M.; Bligaard, T., Active Learning Accelerated Discovery of Stable Iridium Oxide Polymorphs for the Oxygen Evolution Reaction. *Chem. Mater.* **2020,** *32* (13), 5854-5863.
34. Janet, J. P.; Ramesh, S.; Duan, C.; Kulik, H. J., Accurate Multiobjective Design in a Space of Millions of Transition Metal Complexes with Neural-Network-Driven Efficient Global Optimization. *ACS Cent. Sci.* **2020,** *6* (4), 513-524.
35. Nandy, A.; Duan, C.; Goffinet, C.; Kulik, H. J., New Strategies for Direct Methane-to-Methanol Conversion from Active Learning Exploration of 16 Million Catalysts. *ChemRxiv* **2022**, 10.26434/chemrxiv-2022-kg3s5.
36. Zhai, H. C.; Alexandrova, A. N., Ensemble-Average Representation of Pt Clusters in Conditions of Catalysis Accessed through GPU Accelerated Deep Neural Network Fitting Global Optimization. *J. Chem. Theory Comput.* **2016,** *12* (12), 6213-6226.
37. Kleifeld, O.; Frenkel, A.; Martin, J. M. L.; Sagi, I., Active site electronic structure and dynamics during metalloenzyme catalysis. *Nat. Struct. Biol.* **2003,** *10* (2), 98-103.
38. Szécsényi, Á.; Li, G.; Gascon, J.; Pidko, E. A., Mechanistic Complexity of Methane Oxidation with $H_2O_2$ by Single-Site Fe/ZSM-5 Catalyst. *ACS Catal.* **2018,** *8* (9), 7961-7972.
39. Vitillo, J. G.; Bhan, A.; Cramer, C. J.; Lu, C. C.; Gagliardi, L., Quantum Chemical Characterization of Structural Single Fe(II) Sites in MIL-Type Metal–Organic Frameworks for the Oxidation of Methane to Methanol and Ethane to Ethanol. *ACS Catal.* **2019,** *9* (4), 2870-2879.
40. Simm, G. N.; Reiher, M., Error-Controlled Exploration of Chemical Reaction Networks with Gaussian Processes. *J. Chem. Theory Comput.* **2018,** *14* (10), 5238-5248.
41. Steiner, M.; Reiher, M., Autonomous Reaction Network Exploration in Homogeneous and Heterogeneous Catalysis. *Top. Catal.* **2022,** *65* (1-4), 6-39.
42. Baiardi, A.; Grimmel, S. A.; Steiner, M.; Turtscher, P. L.; Unsleber, J. P.; Weymuth, T.; Reiher, M., Expansive Quantum Mechanical Exploration of Chemical Reaction Paths. *Acc. Chem. Res.* **2022,** *55* (1), 35-43.
43. Guan, Y. F.; Ingman, V. M.; Rooks, B. J.; Wheeler, S. E., AARON: An Automated Reaction Optimizer for New Catalysts. *J. Chem. Theory Comput.* **2018,** *14* (10), 5249-5261.
44. Schaefer, A. J.; Ingman, V. M.; Wheeler, S. E., SEQCROW: A ChimeraX bundle to facilitate quantum chemical applications to complex molecular systems. *J. Comput. Chem.* **2021,** *42* (24), 1750-1754.
45. Vaucher, A. C.; Reiher, M., Molecular Propensity as a Driver for Explorative Reactivity Studies. *J. Chem. Inf. Model.* **2016,** *56* (8), 1470-1478.
46. Zassinovich, G.; Mestroni, G.; Gladiali, S., Asymmetric Hydrogen Transfer-Reactions Promoted by Homogeneous Transition-Metal Catalysts. *Chem. Rev.* **1992,** *92* (5), 1051-1069.
47. Durand, D. J.; Fey, N., Computational Ligand Descriptors for Catalyst Design. *Chem. Rev.* **2019,** *119* (11), 6561-6594.
48. de Vries, J. G.; Jackson, S. D., Homogeneous and heterogeneous catalysis in industry. *Catal. Sci. Technol.* **2012,** *2* (10), 2009-2009.
49. Slaugh, L. H.; Mullineaux, R. D., Novel Hydroformylation Catalysts. *J. Organomet. Chem.* **1968,** *13* (2), 469-+.

50. Trnka, T. M.; Grubbs, R. H., The development of L2X2Ru = CHR olefin metathesis catalysts: An organometallic success story. *Acc. Chem. Res.* **2001,** *34* (1), 18-29.
51. Twilton, J.; Le, C.; Zhang, P.; Shaw, M. H.; Evans, R. W.; MacMillan, D. W. C., The merger of transition metal and photocatalysis. *Nat. Rev. Chem.* **2017,** *1* (7).
52. Cheung, K. C.; Wong, W. L.; Ma, D. L.; Lai, T. S.; Wong, K. Y., Transition metal complexes as electrocatalysts - Development and applications in electro-oxidation reactions. *Coordin. Chem. Rev.* **2007,** *251* (17-20), 2367-2385.
53. Shaik, S.; Chen, H.; Janardanan, D., Exchange-Enhanced Reactivity in Bond Activation by Metal-Oxo Enzymes and Synthetic Reagents. *Nat. Chem.* **2011,** *3* (1), 19-27.
54. Duan, C.; Janet, J. P.; Liu, F.; Nandy, A.; Kulik, H. J., Learning from Failure: Predicting Electronic Structure Calculation Outcomes with Machine Learning Models. *J. Chem. Theory Comput.* **2019,** *15* (4), 2331-2345.
55. Nandy, A.; Duan, C.; Janet, J. P.; Gugler, S.; Kulik, H. J., Strategies and Software for Machine Learning Accelerated Discovery in Transition Metal Chemistry. *Ind. Eng. Chem. Res.* **2018,** *57* (42), 13973-13986.
56. Heinen, S.; Schwilk, M.; von Rudorff, G. F.; von Lilienfeld, O. A., Machine learning the computational cost of quantum chemistry. *Mach. Learn.-Sci. Technol.* **2020,** *1* (2).
57. Ma, S.; Ma, Y. J.; Zhang, B. H.; Tian, Y. Q.; Jin, Z., Forecasting System of Computational Time of DFT/TDDFT Calculations under the Multiverse Ansatz via Machine Learning and Cheminformatics. *ACS Omega* **2021,** *6* (3), 2001-2024.
58. McAnanama-Brereton, S.; Waller, M. P., Rational Density Functional Selection Using Game Theory. *J. Chem. Inf. Model.* **2018,** *58* (1), 61-67.
59. Janet, J. P.; Kulik, H. J., Resolving Transition Metal Chemical Space: Feature Selection for Machine Learning and Structure–Property Relationships. *The Journal of Physical Chemistry A* **2017,** *121* (46), 8939-8954.
60. Duan, C.; Liu, F.; Nandy, A.; Kulik, H. J., Putting Density Functional Theory to the Test in Machine-Learning-Accelerated Materials Discovery. *The Journal of Physical Chemistry Letters* **2021,** *12* (19), 4628-4637.
61. Tsai, S. T.; Kuo, E. J.; Tiwary, P., Learning molecular dynamics with simple language model built upon long short-term memory neural network. *Nat. Commun.* **2020,** *11* (1).
62. Wang, Y. H.; Ribeiro, J. M. L.; Tiwary, P., Machine learning approaches for analyzing and enhancing molecular dynamics simulations. *Curr. Opin. Struc. Biol.* **2020,** *61*, 139-145.
63. Kates-Harbeck, J.; Svyatkovskiy, A.; Tang, W., Predicting disruptive instabilities in controlled fusion plasmas through deep learning. *Nature* **2019,** *568* (7753), 526-+.
64. Degrave, J.; Felici, F.; Buchli, J.; Neunert, M.; Tracey, B.; Carpanese, F.; Ewalds, T.; Hafner, R.; Abdolmaleki, A.; de las Casas, D.; Donner, C.; Fritz, L.; Galperti, C.; Huber, A.; Keeling, J.; Tsimpoukelli, M.; Kay, J.; Merle, A.; Moret, J.-M.; Noury, S.; Pesamosca, F.; Pfau, D.; Sauter, O.; Sommariva, C.; Coda, S.; Duval, B.; Fasoli, A.; Kohli, P.; Kavukcuoglu, K.; Hassabis, D.; Riedmiller, M., Magnetic control of tokamak plasmas through deep reinforcement learning. *Nature* **2022,** *602* (7897), 414-419.
65. Olah, G. A., Beyond Oil and Gas: the Methanol Economy. *Angew. Chem., Int. Ed.* **2005,** *44* (18), 2636-2639.
66. Lunsford, J. H., Catalytic Conversion of Methane to More Useful Chemicals and Fuels: a Challenge for the 21st Century. *Catal. Today* **2000,** *63* (2), 165-174.

67. Shiota, Y.; Yoshizawa, K., Methane-to-methanol conversion by first-row transition-metal oxide ions: ScO+TiO+, VO+, CrO+, MnO+, FeO+, CoO+, NiO+, and CuO+. *J. Am. Chem. Soc.* **2000,** *122* (49), 12317-12326.
68. Nandy, A.; Kulik, H. J., Why Conventional Design Rules for C–H Activation Fail for Open-Shell Transition-Metal Catalysts. *ACS Catal.* **2020,** *10* (24), 15033-15047.
69. Huang, X. Y.; Groves, J. T., Beyond ferryl-mediated hydroxylation: 40 years of the rebound mechanism and C-H activation. *J. Biol. Inorg. Chem.* **2017,** *22* (2-3), 185-207.
70. McInnes, L.; Healy, J.; Melville, J., UMAP: Uniform Manifold Approximation and Projection for Dimension Reduction. *arXiv:1802.03426* **2018**.
71. Fukuzumi, S.; Kojima, T.; Lee, Y. M.; Nam, W., High-valent metal-oxo complexes generated in catalytic oxidation reactions using water as an oxygen source. *Coordin. Chem. Rev.* **2017,** *333*, 44-56.
72. Kang, Y.; Chen, H.; Jeong, Y. J.; Lai, W.; Bae, E. H.; Shaik, S.; Nam, W., Enhanced Reactivities of Iron(IV)-Oxo Porphyrin pi-Cation Radicals in Oxygenation Reactions by Electron-Donating Axial Ligands. *Chem-Eur J.* **2009,** *15* (39), 10039-10046.
73. Selvaraju, R. R.; Cogswell, M.; Das, A.; Vedantam, R.; Parikh, D.; Batra, D., Grad-CAM: Visual Explanations from Deep Networks via Gradient-based Localization. *IEEE Int. Conf. Comp. Vis.* **2017**, 618-626.
74. Petachem. TeraChem. http://www.petachem.com/ (accessed May 17, 2019).
75. Seritan, S.; Bannwarth, C.; Fales, B. S.; Hohenstein, E. G.; Isborn, C. M.; Kokkila-Schumacher, S. I. L.; Li, X.; Liu, F.; Luehr, N.; Snyder Jr., J. W.; Song, C.; Titov, A. V.; Ufimtsev, I. S.; Wang, L.-P.; Martínez, T. J., TeraChem: A graphical processing unit-accelerated electronic structure package for large-scale ab initio molecular dynamics. *WIREs Comput. Mol. Sci.* **2021,** *11* (2), e1494.
76. Becke, A. D., Density-Functional Thermochemistry. III. The Role of Exact Exchange. *The Journal of Chemical Physics* **1993,** *98* (7), 5648-5652.
77. Lee, C.; Yang, W.; Parr, R. G., Development of the Colle-Salvetti Correlation-Energy Formula into a Functional of the Electron Density. *Phys. Rev. B* **1988,** *37*, 785-789.
78. Stephens, P. J.; Devlin, F. J.; Chabalowski, C. F.; Frisch, M. J., Ab Initio Calculation of Vibrational Absorption and Circular Dichroism Spectra Using Density Functional Force Fields. *J. Phys. Chem.* **1994,** *98* (45), 11623-11627.
79. Grimme, S.; Antony, J.; Ehrlich, S.; Krieg, H., A Consistent and Accurate Ab Initio Parametrization of Density Functional Dispersion Correction (DFT-D) for the 94 Elements H-Pu. *The Journal of Chemical Physics* **2010,** *132* (15), 154104.
80. Becke, A. D.; Johnson, E. R., A Density-Functional Model of the Dispersion Interaction. *The Journal of Chemical Physics* **2005,** *123* (15), 154101.
81. Wadt, W. R.; Hay, P. J., Ab Initio Effective Core Potentials for Molecular Calculations. Potentials for Main Group Elements Na to Bi. *The Journal of Chemical Physics* **1985,** *82* (1), 284-298.
82. Hay, P. J.; Wadt, W. R., Ab Initio Effective Core Potentials for Molecular Calculations. Potentials for the Transition Metal Atoms Sc to Hg. *The Journal of Chemical Physics* **1985,** *82* (1), 270-283.
83. Rassolov, V. A.; Pople, J. A.; Ratner, M. A.; Windus, T. L., 6-31G* Basis Set for Atoms K through Zn. *The Journal of Chemical Physics* **1998,** *109* (4), 1223-1229.
84. Saunders, V. R.; Hillier, I. H., A "Level-Shifting" Method for Converging Closed Shell Hartree-Fock Wave Functions. *Int. J. Quantum Chem.* **1973,** *7* (4), 699-705.

85. Wang, L.-P.; Song, C., Geometry Optimization Made Simple with Translation and Rotation Coordinates. *The Journal of Chemical Physics* **2016,** *144* (21), 214108.
86. Ioannidis, E. I.; Gani, T. Z. H.; Kulik, H. J., molSimplify: A Toolkit for Automating Discovery in Inorganic Chemistry. *J. Comput. Chem.* **2016,** *37*, 2106-2117.
87. Group, K. molSimplify & molSimplify Automatic Design. https://github.com/hjkgrp/molsimplify (accessed June 24, 2021).
88. Groves, J. T.; McClusky, G. A., Aliphatic Hydroxylation via Oxygen Rebound. Oxygen Transfer Catalyzed by Iron. *J. Am. Chem. Soc.* **1976,** *98* (3), 859-861.
89. O'Boyle, N. M.; Banck, M.; James, C. A.; Morley, C.; Vandermeersch, T.; Hutchison, G. R., Open Babel: An Open Chemical Toolbox. *J. Cheminf.* **2011,** *3*, 33.
90. O'Boyle, N. M.; Morley, C.; Hutchison, G. R., Pybel: a Python Wrapper for the OpenBabel Cheminformatics Toolkit. *Chem. Cent. J.* **2008,** *2*, 5.
91. Duan, C.; Liu, F.; Nandy, A.; Kulik, H. J., Putting Density Functional Theory to the Test in Machine-Learning-Accelerated Materials Discovery. *J. Phys. Chem. Lett.* **2021,** *12* (19), 4628-4637.
92. Chollet, F. Keras. https://keras.io/ (accessed June 24, 2021).
93. Abadi, M.; Agarwal, A.; Barham, P.; Brevdo, E.; Chen, Z.; Citro, C.; Corrado, G. S.; Davis, A.; Dean, J.; Devin, M.; Ghemawat, S.; Goodfellow, I.; Harp, A.; Irving, G.; Isard, M.; Jozefowicz, R.; Jia, Y.; Kaiser, L.; Kudlur, M.; Levenberg, J.; Mané, D.; Schuster, M.; Monga, R.; Moore, S.; Murray, D.; Olah, C.; Shlens, J.; Steiner, B.; Sutskever, I.; Talwar, K.; Tucker, P.; Vanhoucke, V.; Vasudevan, V.; Viégas, F.; Vinyals, O.; Warden, P.; Wattenberg, M.; Wicke, M.; Yu, Y.; Zheng, X. *TensorFlow: Large-Scale Machine Learning on Heterogeneous Systems*, 2015.

**Electronic Supplementary Information for**

***Machine learning models predict calculation outcomes with the transferability necessary for computational catalysis***

Chenru Duan[1,2], Aditya Nandy[1,2], Husain Adamji[1], Yuriy Roman-Leshkov[1], and Heather J. Kulik[1,*]

[1]Department of Chemical Engineering, Massachusetts Institute of Technology, Cambridge, MA 02139

[2]Department of Chemistry, Massachusetts Institute of Technology, Cambridge, MA 02139

*email: hjkulik@mit.edu

**Contents**

**Table S1.** Geometry cutoffs used in this work. Geometry check metrics are the same as our prior work on transition metal–oxo catalysts.[1] Strict cutoffs applied on a completed optimization are indicated alongside loose cutoffs applied during each resubmission in parentheses. These geometry checks include: coordination number, which must be 6; mean and max. deviation of connecting-atom-metal-connecting atom angles; max. deviation of any metal–ligand bond length; max. deviation of metal–ligand bond lengths for equatorial ligands; max. root mean square deviation (RMSD) of the ligand from its initial structure; and mean and max. deviation of an expected linear ligand from a 180° angle (which are only computed for linear ligands). These checks are applied for all intermediates that are geometry optimized.

| **coordination number** | | | |
|---|---|---|---|
| 6 (6) | | | |
| **first coordination sphere shape** | | | |
| mean(Δθ($C_i$-M-$C_j$)) | max(Δθ($C_i$-M-$C_j$)) | max(Δd) | max($\Delta d_{eq}$) |
| 30° (35°) | 90° (90°) | 1.6 Å (1.6 Å) | 0.55 Å (0.55 Å) |
| **ligand distortion metrics** | | | |
| max(RMSD) | | mean(Δθ(M-A-B)) | max(Δθ(M-A-B)) |
| 3.0 Å (4.0 Å) | | 20° (30°) | 28° (40°) |

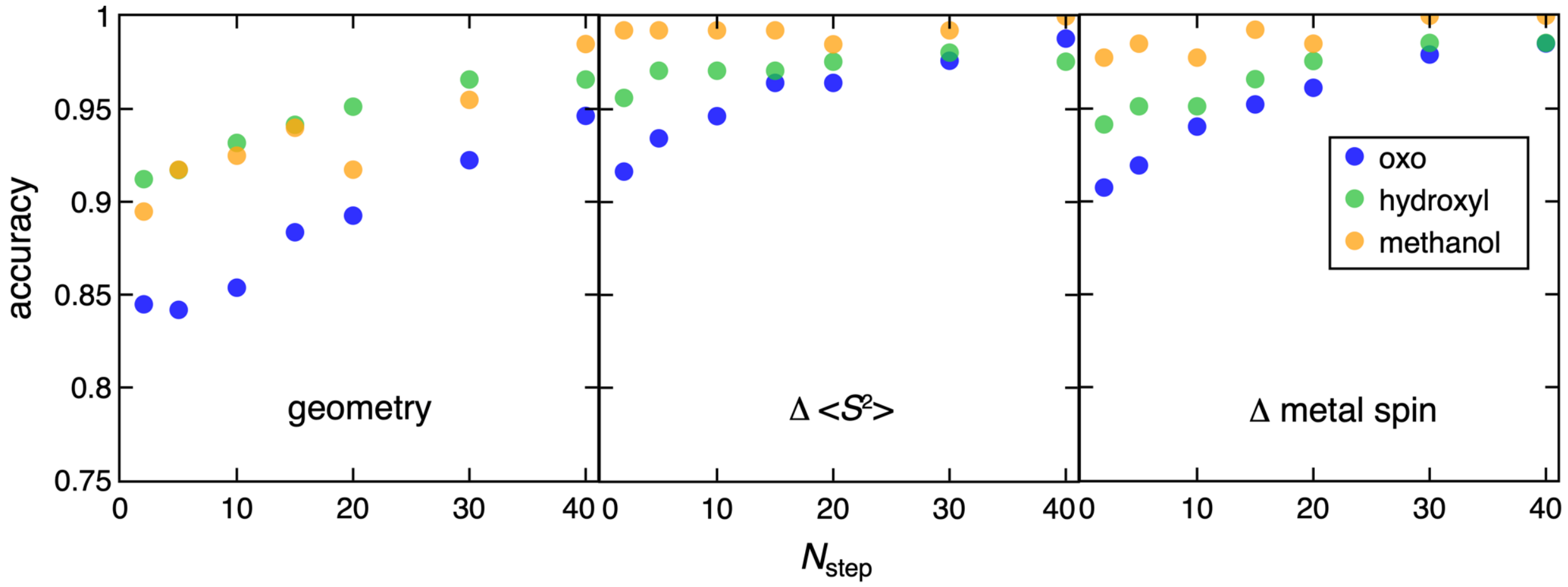

**Figure S1.** Model accuracy versus the number of geometry optimization steps for the dynamic classifiers at each $N_{step}$ (i.e., 2, 5, 10, 15, 20, 30, and 40) evaluated on the set-aside test set of the *WC* set grouped by intermediate (blue for metal–oxo, green for metal–hydroxo, and orange for methanol–bound). The performance of each task: geometry (left), ⟨$S^2$⟩ deviation (middle), and metal spin deviation (right), is reported separately. The dynamic classifiers are trained on all three intermediates in the *WC* set. Note that the models make predictions on all set-aside test points at all steps (i.e., data fraction is always 1 and no LSE cutoff is used).

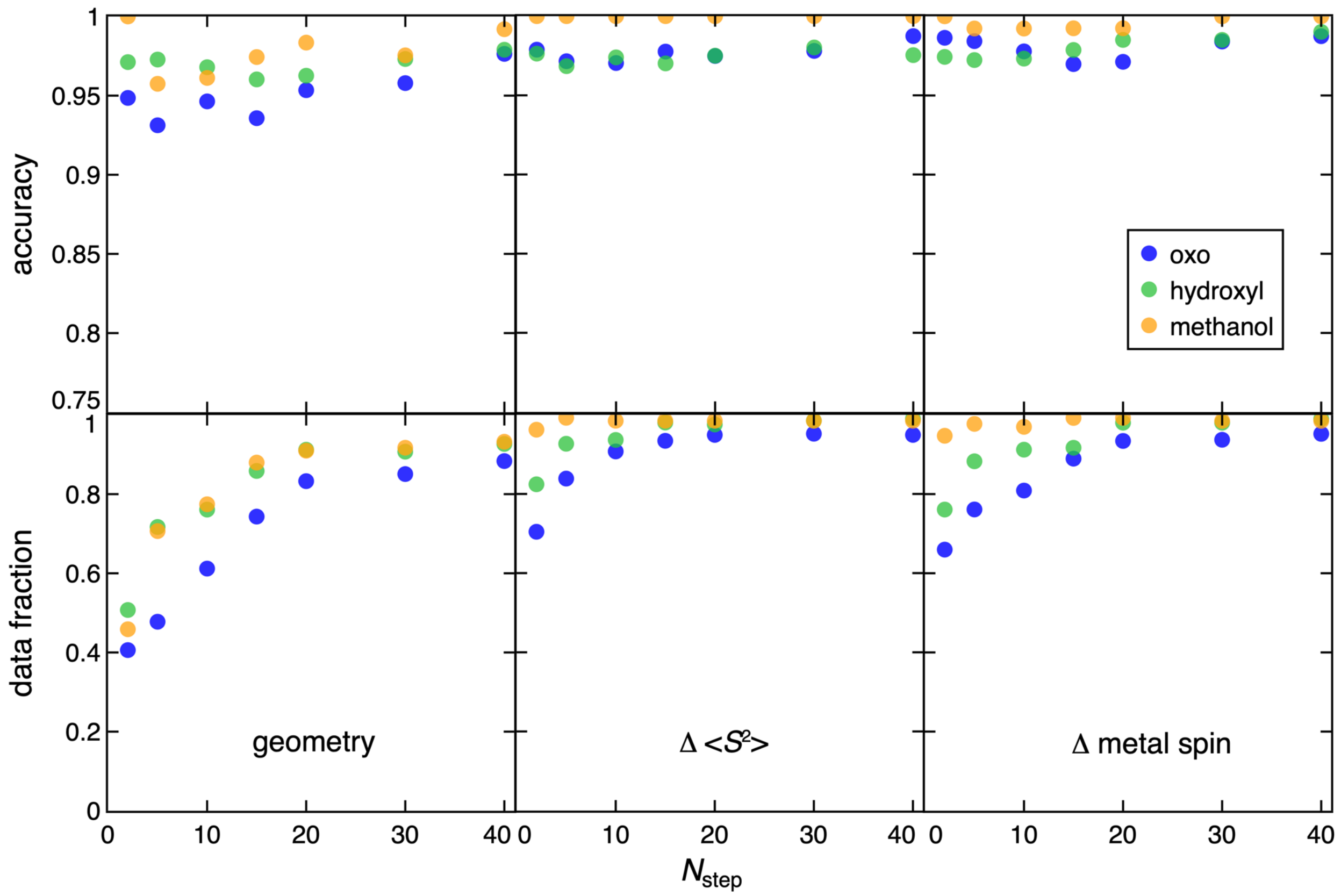

**Figure S2.** Uncertainty-aware model accuracy (top) and data fraction (bottom) versus the number of geometry optimization steps with a LSE cutoff at 0.3 for the dynamic classifiers at each $N_{\text{step}}$ (i.e., 2, 5, 10, 15, 20, 30, and 40) evaluated on the set-aside test set of the *WC* set grouped by intermediate (blue for metal–oxo, green for metal–hydroxo, and orange for methanol–bound). The performance of each task: geometry (left), $\langle S^2 \rangle$ deviation (middle), and metal spin deviation (right), is reported separately. The dynamic classifiers are trained on all three intermediates in the *WC* set.

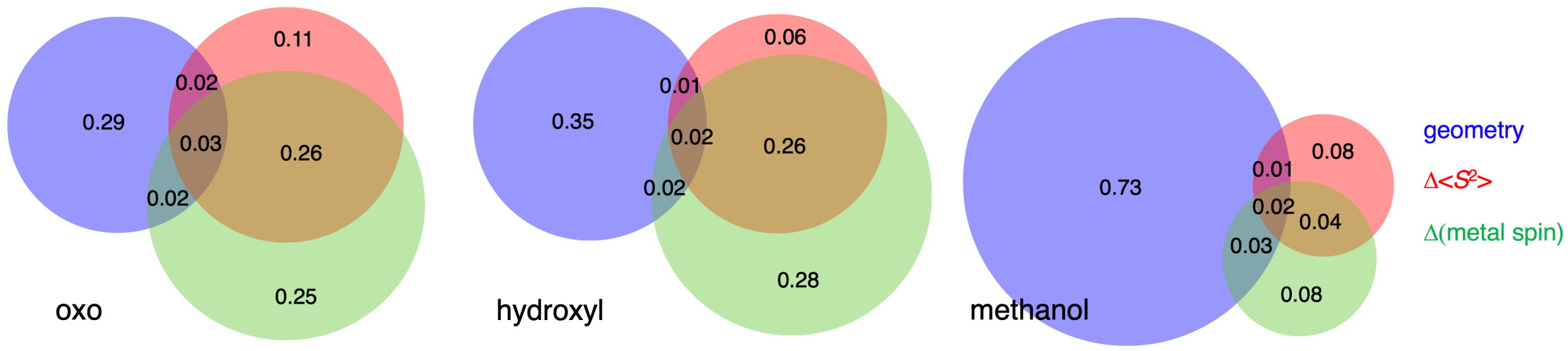

**Figure S3.** Venn diagrams for fractions of failure modes (blue for geometry, red for $\langle S^2 \rangle$ deviation, and green for metal spin deviation) of three intermediates: metal–oxo (left), metal–hydroxo (middle), and methanol–bound (right) in the *WC* set. The distribution of failure modes is distinct for three intermediates and that the concurrence of $\langle S^2 \rangle$ and metal spin deviation is high.

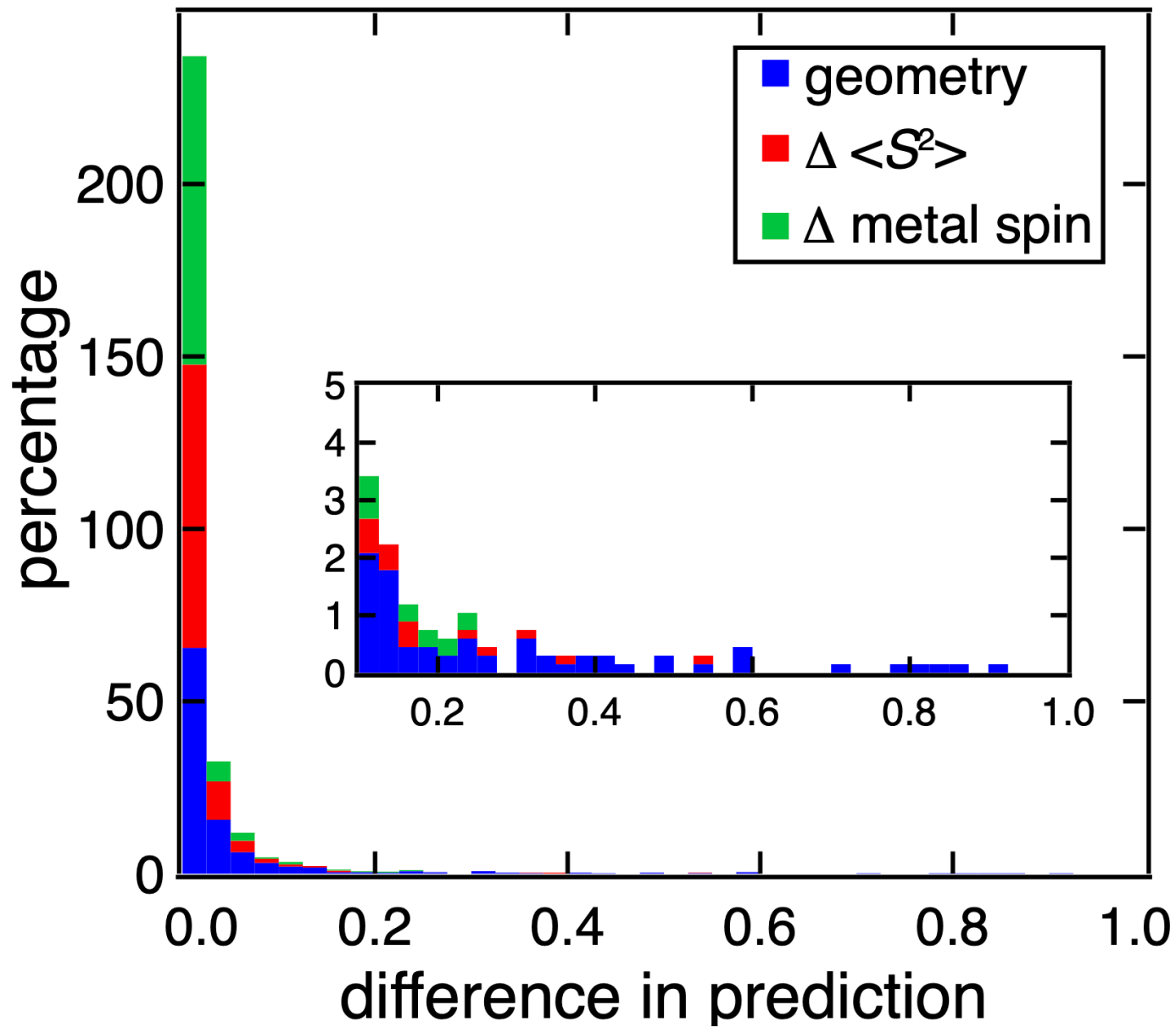

**Figure S4.** Stacked histogram for absolute difference in predictions between the 40-step dynamic classifier trained on all three intermediates in the *WC* set and the 40-step dynamic classifier trained on only the metal–oxo intermediate in the *WC* set. All three predictions tasks are shown: blue for geometry, red for $\langle S^2 \rangle$ deviation, and green for metal spin deviation. A zoomed-in view for the percentage of cases where the deviation between two dynamic classifiers is > 0.1 is shown as an inset. The output of each dynamic classifier is a probability from 0 (bad) to 1 (good). The difference is evaluated on the set-aside test set of the *WC* set with a bin size 0.025. Since histograms of three prediction tasks are stacked together, the total percentage should sum up to 300 across all bins.

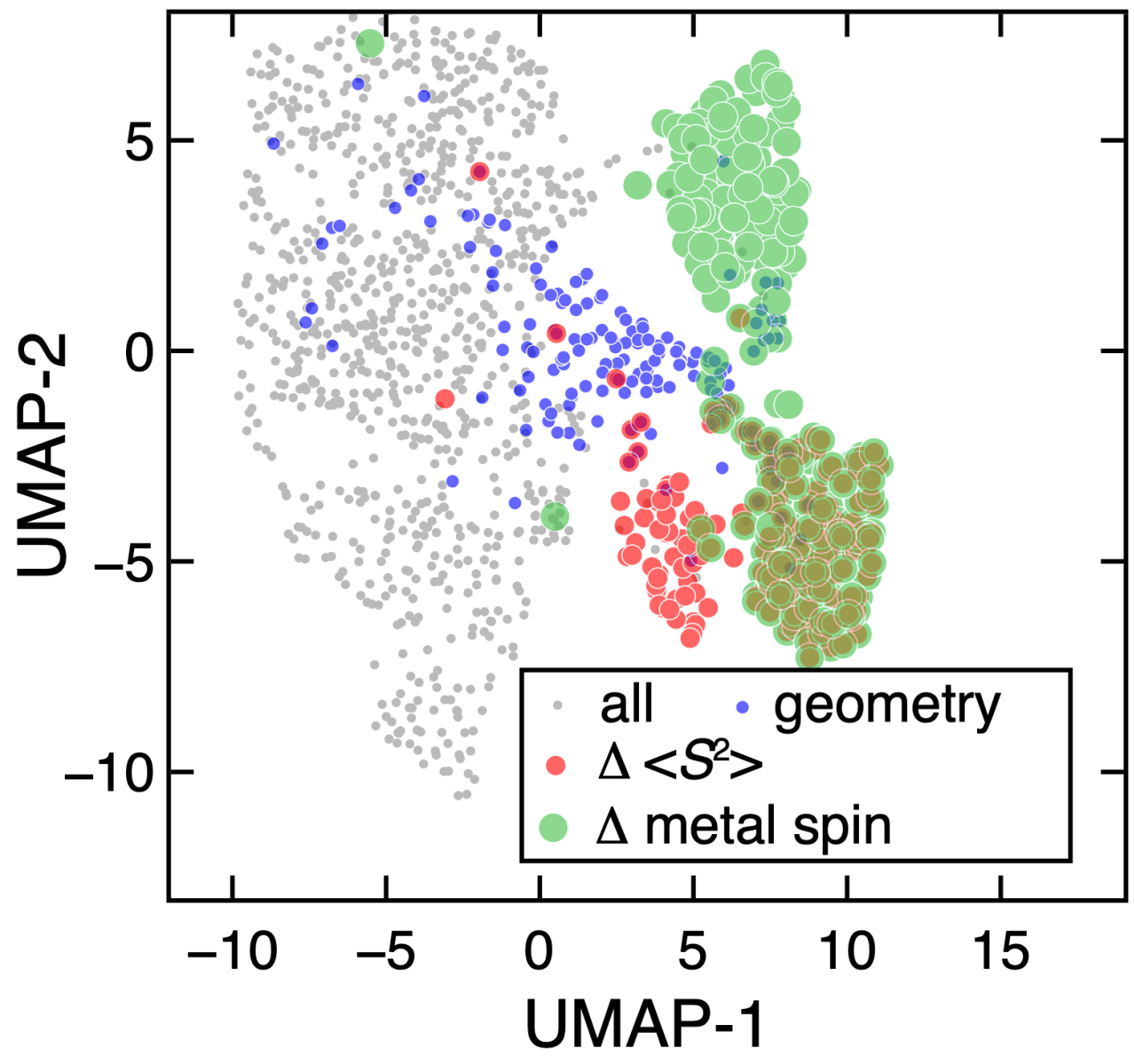

**Figure S5.** UMAP[5] 2D visualization of the latent space of the multi-task dynamic classifier trained on 40 steps of geometry optimization trajectories of only the metal–oxo intermediate in the *WC* set. All data points from the *WC* set are shown in gray. Geometry optimizations that are labeled as bad are colored separately for each reason: blue for geometry, red for $\langle S^2 \rangle$ deviation, and green for metal spin deviation. Different circle sized are used to show more clearly when points overlap. Note that the distributions of the clusters are very similar to Figure 3 in the main text, indicating similar model weights in two sets of dynamic classifiers.

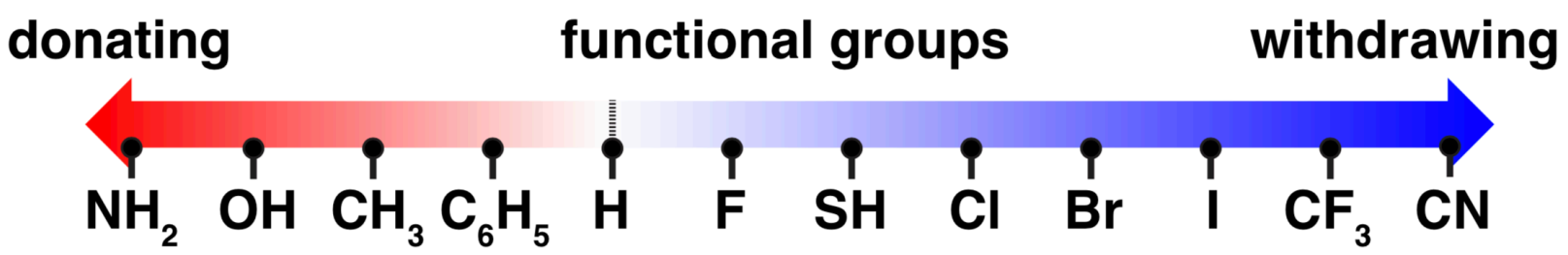

**Figure S6.** Functional groups considered in the *FWC* set with respect to the electron donating/withdrawing effect. This figure is adapted from Figure 5 of our prior work[2].

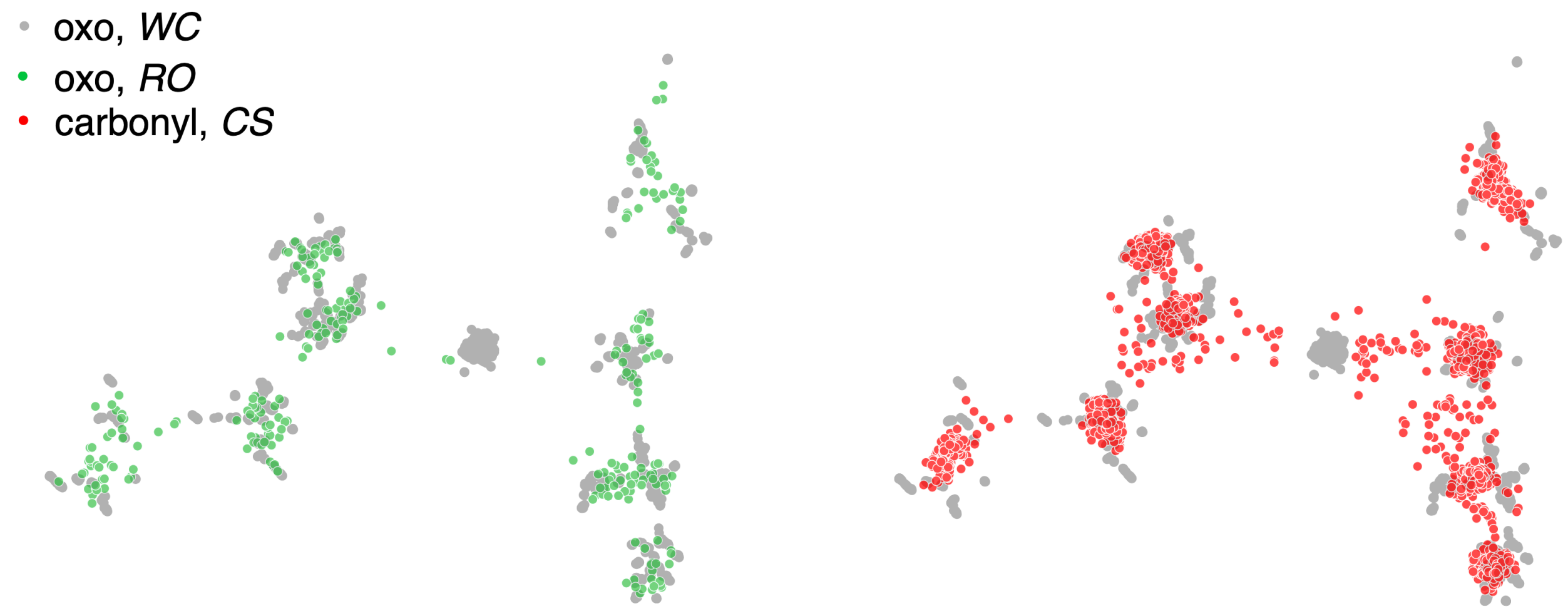

**Figure S7.** UMAP 2D visualization of the RAC-155 features[6] for different intermediates: metal–oxo intermediate in the *WC* set (gray, in both panes), Ru–oxo intermediate in the *RO* set (green, left), and metal–carbonyl intermediate in the *CS* set (red, right).

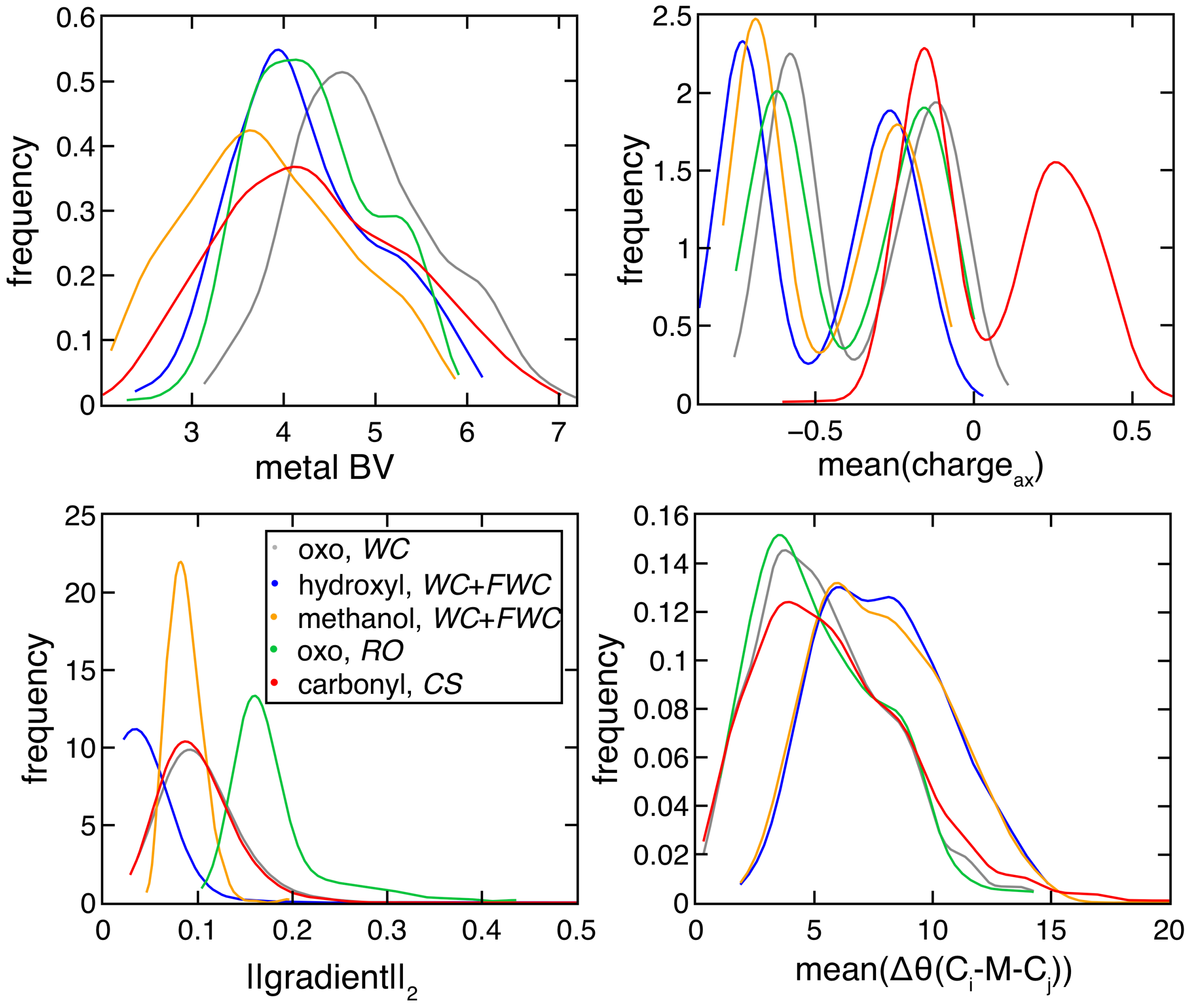

**Figure S8.** Distributions of select initial (i.e., at step 0) electronic structure- and geometry-based features for different intermediates: metal–oxo intermediate in the *WC* set (gray), metal–hydroxo intermediate in the *WC* and *FWC* set (blue), metal–methanol intermediate in the *WC* and *FWC* set (orange), Ru–oxo intermediate in the *RO* set (green), and metal–carbonyl intermediate in the *CS* set (red). Four different features are shown separately: metal bond valence (top left), average Mulliken charge for axial ligands (top right), norm of the energy gradient (bottom left), and average angular deviation of the first coordination shell (bottom right).

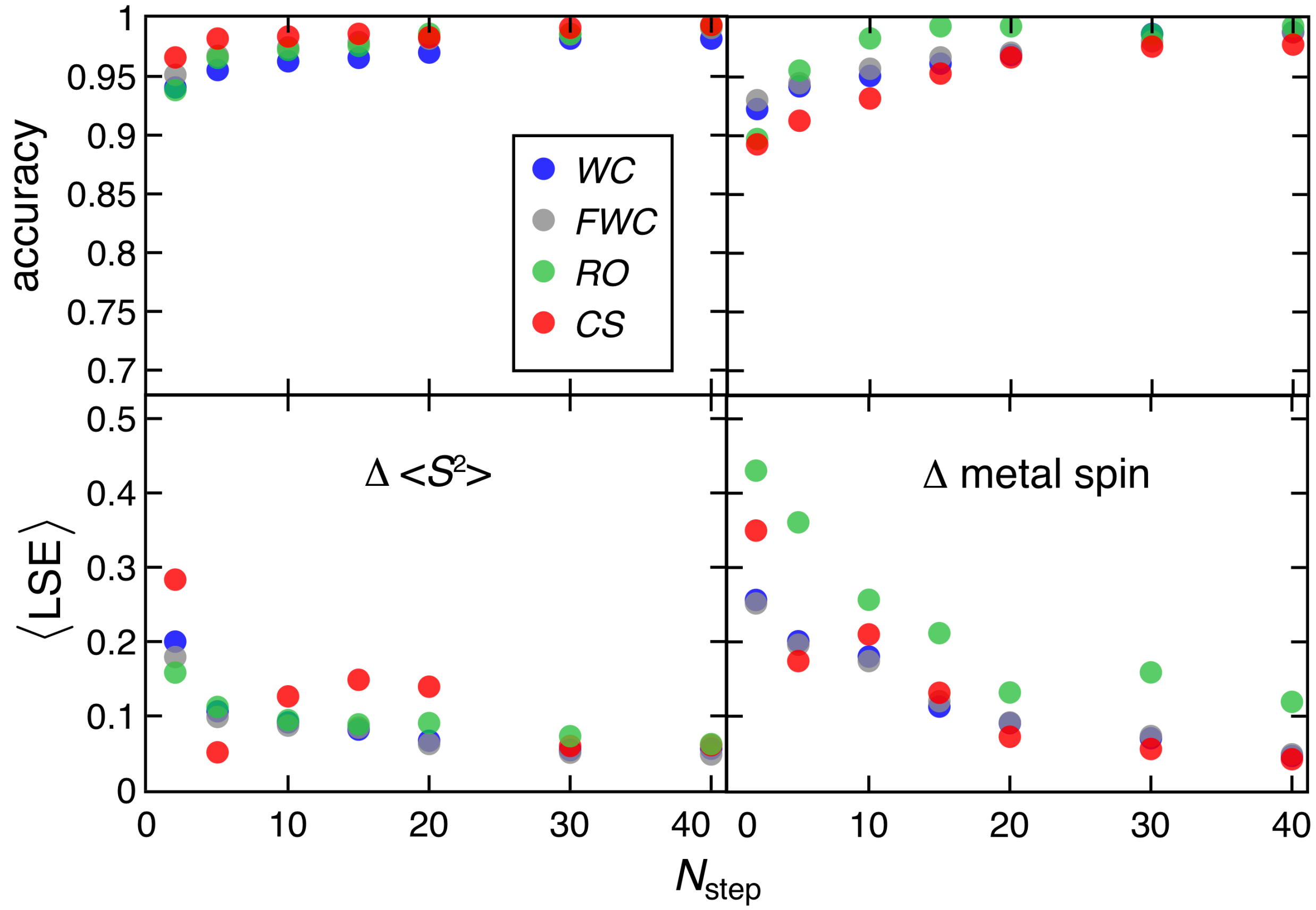

**Figure S9.** Accuracy (top) and the average LSE (bottom) for the ⟨$S^2$⟩ deviation classification (left) and metal spin deviation classification (right) for the set-aside test set in *WC* set (blue) and three out-of-distribution test sets (*FWC* in gray, *RO* in green, and *CS* in red) with increasing number of optimization steps, $N_{step}$. The dynamic classifier was trained only on the metal–oxo intermediate in the *WC* set. Here, the accuracy is reported on all data (i.e., no LSE uncertainty control is used).

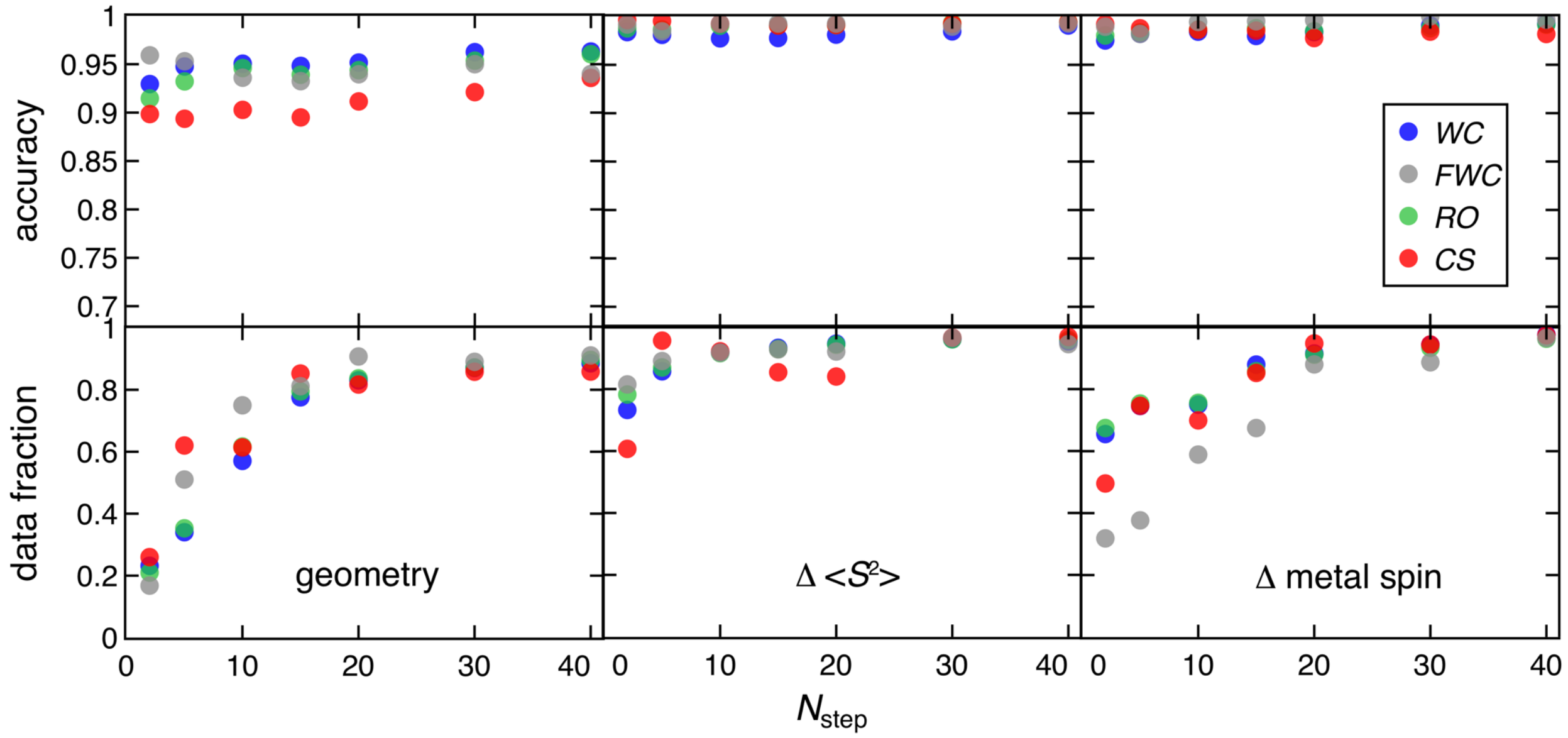

**Figure S10.** Accuracy (top) and data fraction (bottom) for three classification tasks (for the set-aside test set in *WC* set (blue) and three out-of-distribution test sets (*FWC* in gray, *RO* in green, and *CS* in red) with increasing number of optimization steps, $N_{step}$. The performance of each task: geometry (left), $\langle S^2 \rangle$ deviation (middle), and metal spin deviation (right), is reported separately. The dynamic classifier was trained only on the metal–oxo intermediate in the *WC* set. An LSE cutoff of 0.3 is used.

**Table S2.** Electronic structure cutoffs used in this work. Metrics for classifying good and bad electronic structure metrics are the same as prior work on transition metal–oxo catalysts.[1] All geometry optimizations and single point energy calculations are required to pass these criteria to contribute to a reaction energy. For the metal–carbonyl species, only the spin density on the metal is considered due to the absence of the connecting oxo.

| **electronic structure check** | **cutoff** |
|---|---|
| $\langle S^2 \rangle - S(S+1)$ | 1 $\mu_B^2$ |
| metal spin density + metal–oxo spin density – total spin density | 1 $\mu_B$ |

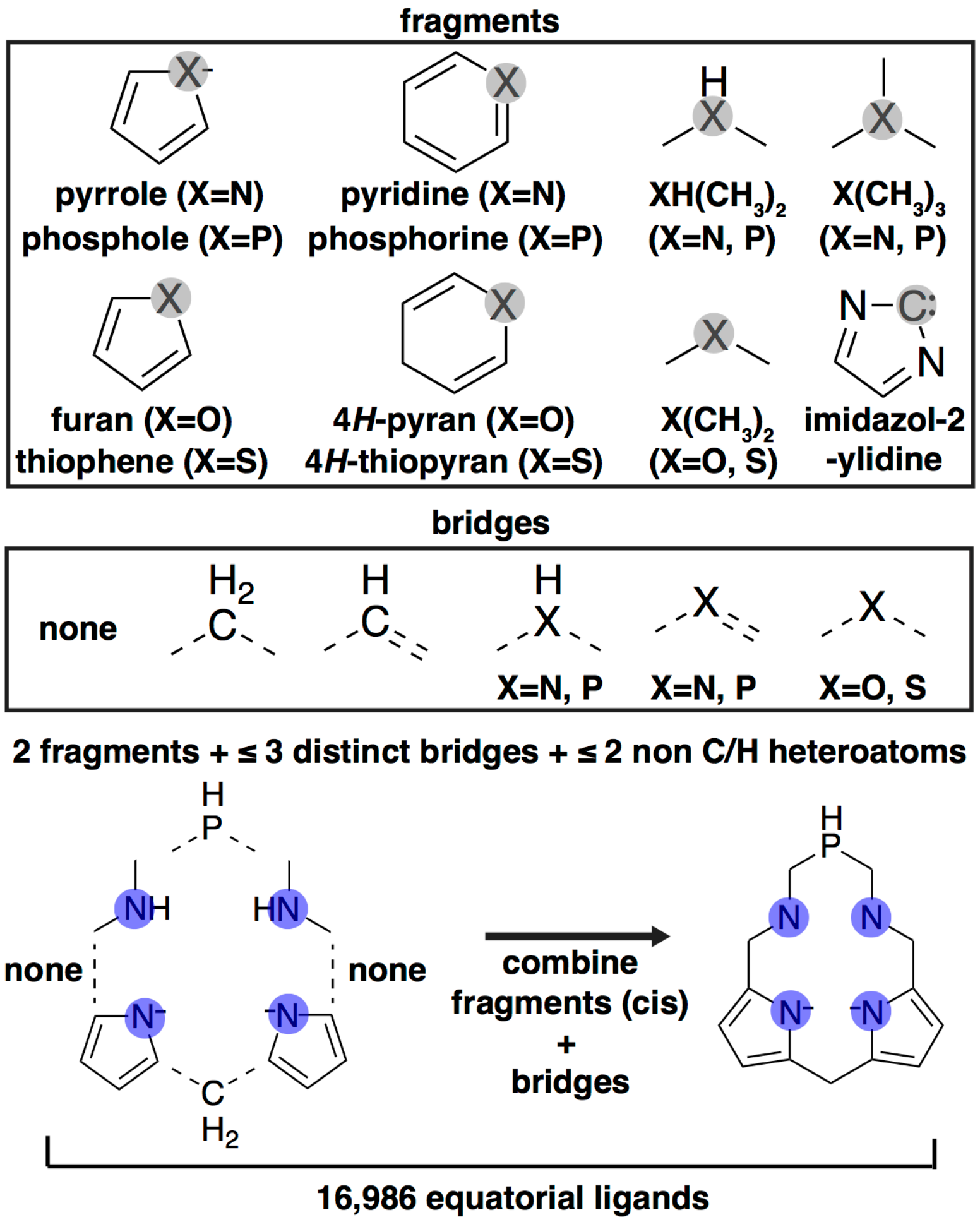

**Figure S11.** Strategies of constructing tetradentate macrocycles used in our previous work[2] using 15 fragments (top) and 9 bridges (middle). All metal-coordinating atoms are highlighted by gray circles, and X is used to indicate the possibility of multiple metal-coordinating element types. Two fragment types in *cis* orientation are combined with up to three distinct bridges to construct a macrocycle. (bottom) An example macrocycle is shown that is constructed from two dimethylamine fragments and two pyrrole fragments joined via one phosphorus bridge, one methylene bridge, and two bridges with no atoms. This Figure is adapted from our prior work[2].

**Table S3**. Formal oxidation and spin states considered in this work for each of the catalytic intermediates studied. We simulate metal–hydroxo intermediates with ferromagnetically coupled hydrogen atom transfer to the metal–oxo intermediate. Consequently, we simulate methanol bound intermediates with antiferromagnetically coupled methyl group addition. Therefore, we simulate the spin-conserved energy landscape, conserving spin between resting state, metal-oxo, and methanol-bound intermediates, and have an additional majority spin hydrogen atom for metal-hydroxo intermediates. For Ru complexes, we only considered Ru–oxo intermediate.

| | formal oxidation state | | | | | spin states | | | | |
|---|---|---|---|---|---|---|---|---|---|---|
| metal | resting state | metal-oxo | metal-hydroxo | methanol-bound | metal-carbonyl | resting state | metal-oxo | metal-hydroxo | methanol-bound | metal-carbonyl |
| Mn | 2 | 4 | 3 | 2 | 2 | 2, 4 | 2, 4 | 3, 5 | 2, 4 | 2, 4, 6 |
| Mn | 3 | 5 | 4 | 3 | 3 | 1, 3 | 1, 3 | 2, 4 | 1, 3 | 1, 3 |
| Fe | 2 | 4 | 3 | 2 | 2 | 1, 3, 5 | 1, 3, 5 | 2, 4, 6 | 1, 3, 5 | 1, 3, 5 |
| Fe | 3 | 5 | 4 | 3 | 3 | 2, 4 | 2, 4 | 3, 5 | 2, 4 | 2, 4 |
| Ru | -- | 4 | -- | -- | -- | -- | 1, 3, 5 | -- | -- | -- |
| Ru | -- | 5 | -- | -- | -- | -- | 2, 4 | -- | -- | -- |

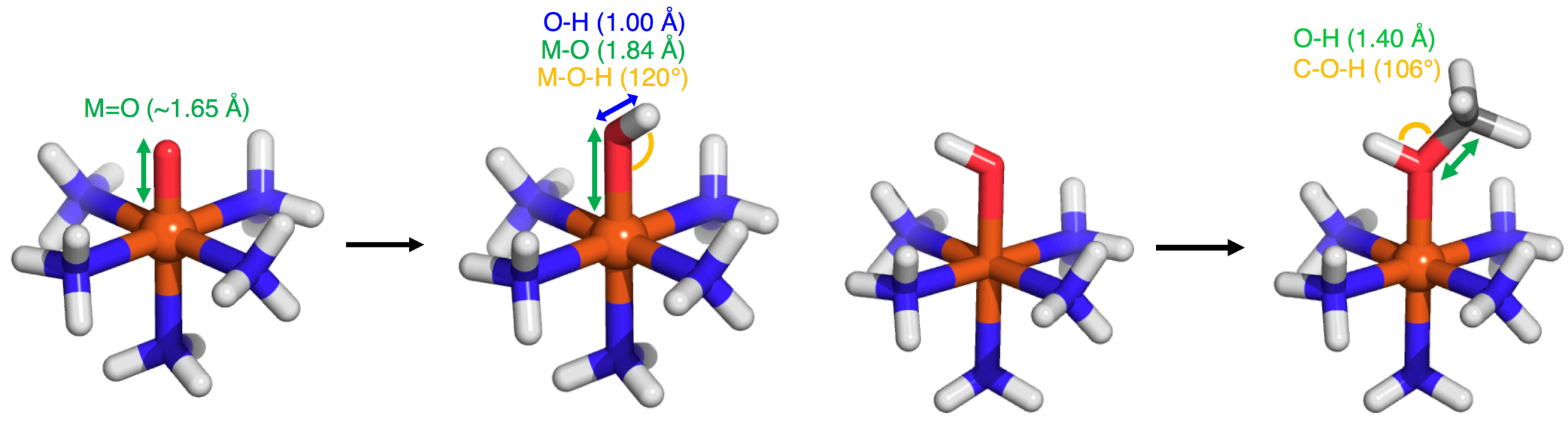

**Figure S12.** Approach for generating a metal–hydroxo geometry from a geometry-optimized metal–oxo structure (left) and approach for generating a methanol bound geometry from a geometry optimized hydroxyl structure (right). For all metal–oxo geometries, the M–O bond was stretched to 1.84 Å, and the hydrogen atom was placed at an M–O–H angle of 120°, 1.00 Å from the oxygen atom. This was the initial geometry for all metal–hydroxo structures, which were subsequently optimized. For the methanol intermediate, the C–O–H angle was set to 106° in the initial geometry, and the C–O bond was set to 1.4 Å, based on the structure of a DFT-optimized methyl group of a methanol molecule. This Figure is adapted from the Supporting Information of our prior work[2].

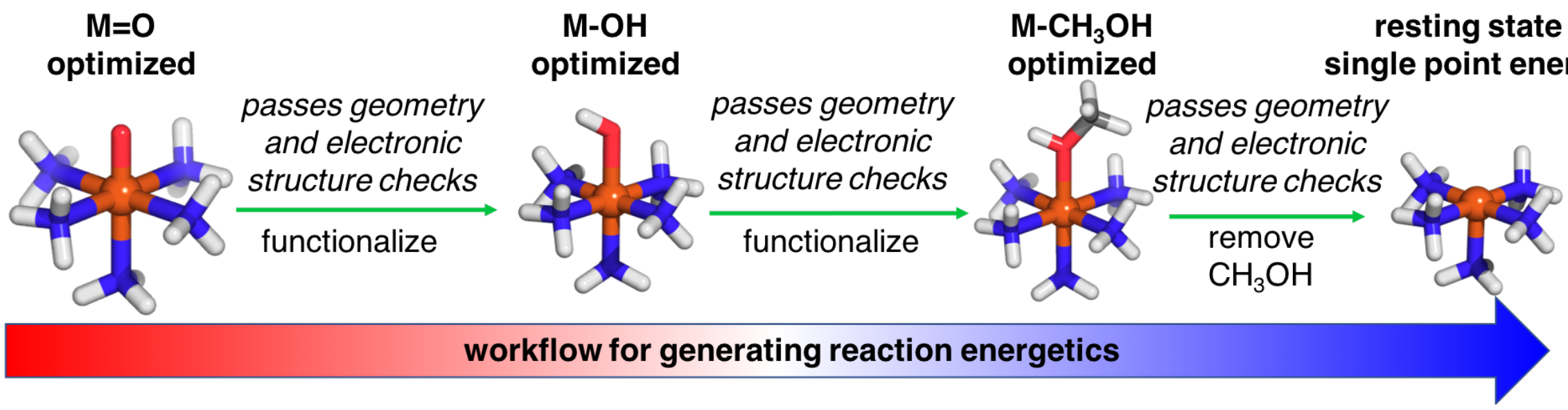

**Figure S13.** Workflow for obtaining reaction energetics. The metal–oxo, metal–hydroxo, and metal–methanol intermediates were geometry optimized without constraints, and the resting state was simulated with a single point energy calculation. After each intermediate, the calculation is checked for fidelity. Only structures that pass all checks move on to the next intermediate. This Figure is adapted from the Supporting Information of our prior work[2].

**Table S4.** Job statistics for metal–oxo, metal–hydroxo, and metal–methanol intermediates in the *WC* data set.

| **intermediate** | **attempted** | **success** | **success rate** |
|---|---|---|---|
| metal–oxo | 1653 | 1058 | 64.0% |
| metal–hydroxo | 1058 | 652 | 61.6% |
| metal–methanol | 652 | 544 | 83.4% |
| overall | 1653 | 544 | 32.9% |

**Table S5.** Job statistics for metal–oxo, metal–hydroxo, and metal–methanol intermediates in the *FWC* data set, metal–oxo intermediates in *RO* data set, and the metal–carbonyl intermediates in the *CS* data set.

| **data set** | **intermediate** | **attempted** | **success** | **success rate** |
|---|---|---|---|---|
| *FWC* | metal–oxo | 1708 | 1077 | 63.1% |
| | metal–hydroxo | 1077 | 729 | 67.7% |
| | metal–methanol | 729 | 618 | 84.7% |
| | overall | 1708 | 618 | 36.2% |
| *RO* | metal–oxo | 292 | 96 | 32.9% |
| *CS* | metal–carbonyl | 2073 | 1310 | 63.2% |

**Table S6.** A complete list of the 26 electronic structure features[3] plus 2 additional spin density-based features[4] for <ΔS²> and metal-spin deviation (for a total of 28 electronic structure descriptors) and the 7 geometry metric features[3] computed for the dynamic classifier model.

| | | | | | |
|---|---|---|---|---|---|
| **from Mulliken charge** | | | | | |
| $C_{metal}$ | $C_{eqcon}^{avrg}$ | $C_{axcon}^{avrg}$ | | | |
| **from bond order matrix** | | | | | |
| $BV_{metal}$ | $BO_{eqcon}^{avrg}$ | $BO_{axcon}^{avrg}$ | $SV_1^{BO}$ | $SV_2^{BO}$ | $SV_3^{BO}$ |
| $SV_4^{BO}$ | $SV_1^{BO,diag=0}$ | $SV_2^{BO,diag=0}$ | $SV_3^{BO,diag=0}$ | $SV_4^{BO,diag=0}$ | |
| **from gradient matrix** | | | | | |
| $G_{metal}$ | $G_{eqcon}^{avrg}$ | $G_{axcon}^{avrg}$ | $G_{max}$ | $G_{RMS}$ | $SV_1^{Grad}$ |
| $SV_2^{Grad}$ | $SV_3^{Grad}$ | $G_{max}^{Grad,mc}$ | $SV_1^{Grad,mc}$ | $SV_2^{Grad,mc}$ | $SV_3^{Grad,mc}$ |
| **from spin density** | | | | | |
| $\langle \Delta S^2 \rangle$ | Δ(metal spin) | | | | |
| **geometric metrics** | | | | | |
| mean(Δθ($C_i$-M-$C_j$)) | max(Δθ($C_i$-M-$C_j$)) | max(Δd) | max($\Delta d_{eq}$) | mean(Δθ(M-A-B)) | max(Δθ(M-A-B)) |
| max(Δθ(M-A-B)) | | | | | |

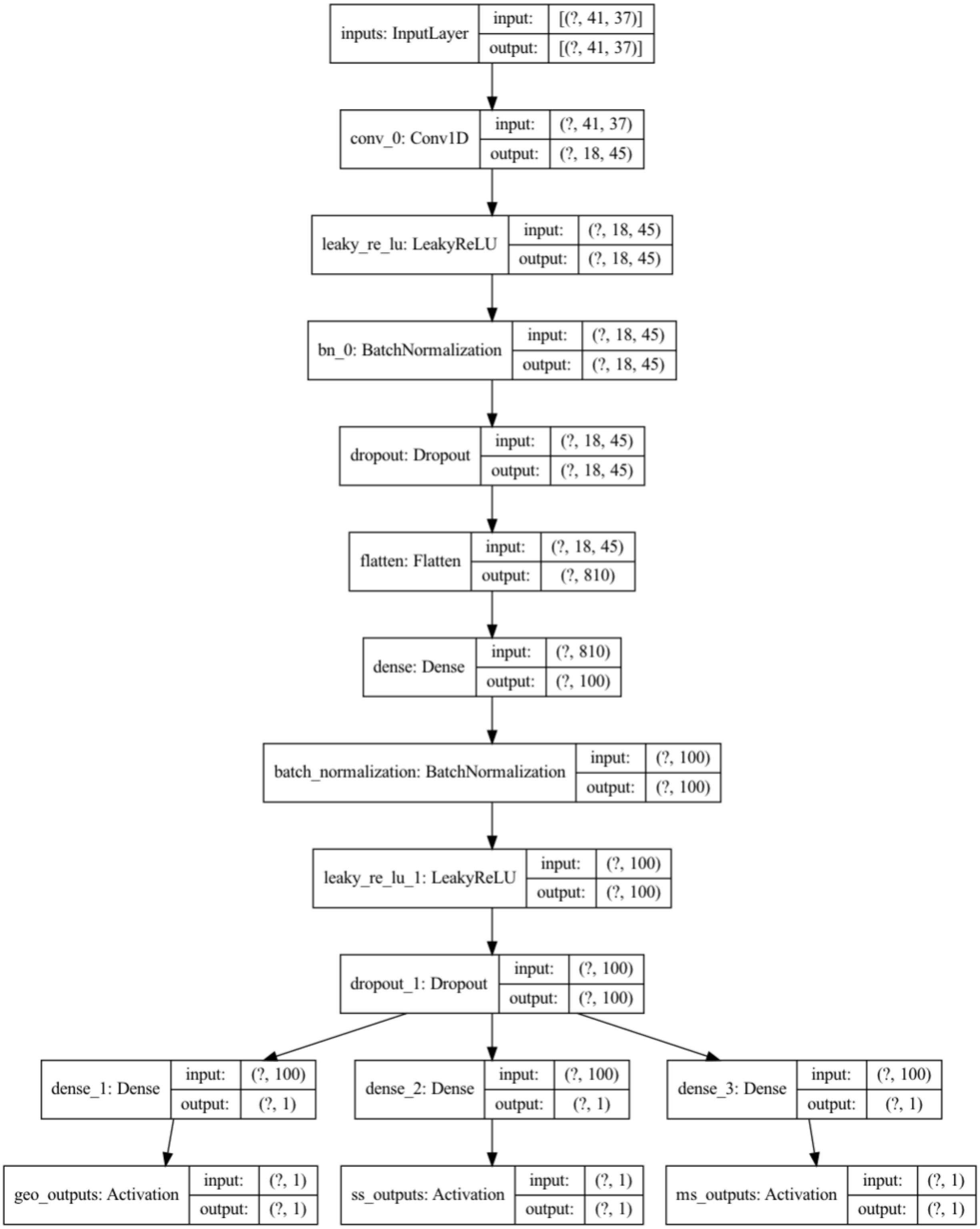

**Figure S14.** Architecture of the multi-task dynamic classifier composed of convolutional and fully-connected layer for prediction of calculation outcomes described in this work. The "?" in the schematic is the size of a batch, which can be flexible and were treated as a hyperparameter.

**Table S7.** A summary of hyperparameters that are used in the dynamic classifiers for the geometry classification task trained on different numbers of steps of geometry optimization.[3] Note that L2-regularization is used in both the convolutional and the dense layer.

| **step of optimization** | **2** | **5** | **10** | **15** | **20** | **30** | **40** |
|---|---|---|---|---|---|---|---|
| **kernel size** | 2 | 3 | 4 | 4 | 5 | 6 | 6 |
| **stride** | 1 | 1 | 1 | 1 | 1 | 2 | 2 |
| **filter** | 35 | 35 | 35 | 35 | 45 | 45 | 45 |
| **dropout rate (convolutional layer)** | 0.3 | 0.3 | 0.3 | 0.3 | 0.3 | 0.3 | 0.3 |
| **dense neurons** | 50 | 50 | 50 | 50 | 50 | 100 | 100 |
| **dropout rate (dense layer)** | 0.1 | 0.1 | 0.1 | 0.1 | 0.1 | 0.2 | 0.2 |
| **L2 regularization** | 1e-4 | 1e-4 | 1e-4 | 1e-4 | 1e-4 | 1e-4 | 1e-4 |
| **learning rate** | 2e-5 | 2e-5 | 2e-5 | 2e-5 | 2e-5 | 1.5e-5 | 1.5e-5 |
| **decay** | 1e-5 | 1e-5 | 1e-5 | 1e-5 | 1e-5 | 1e-5 | 1e-5 |
| **batch size** | 256 | 256 | 256 | 256 | 256 | 256 | 256 |

## References

1. Nandy, A.; Kulik, H. J., Why Conventional Design Rules for C–H Activation Fail for Open-Shell Transition-Metal Catalysts. *Acs Catal* **2020,** *10* (24), 15033-15047.
2. Nandy, A.; Duan, C.; Goffinet, C.; Kulik, H. J., *in preparation* **2021**.
3. Duan, C.; Janet, J. P.; Liu, F.; Nandy, A.; Kulik, H. J., Learning from Failure: Predicting Electronic Structure Calculation Outcomes with Machine Learning Models. *J Chem Theory Comput* **2019,** *15* (4), 2331-2345.
4. Duan, C.; Liu, F.; Nandy, A.; Kulik, H. J., Putting Density Functional Theory to the Test in Machine-Learning-Accelerated Materials Discovery. *The Journal of Physical Chemistry Letters* **2021,** *12* (19), 4628-4637.
5. McInnes, L.; Healy, J.; Melville, J., UMAP: Uniform Manifold Approximation and Projection for Dimension Reduction. *arXiv:1802.03426* **2018**.
6. Janet, J. P.; Kulik, H. J., Resolving Transition Metal Chemical Space: Feature Selection for Machine Learning and Structure–Property Relationships. *The Journal of Physical Chemistry A* **2017,** *121* (46), 8939-8954.